\documentclass[11pt]{article}
\usepackage{bm,graphicx}
\usepackage[english]{babel}
\usepackage{amssymb, amsthm, amsmath}

\newcommand{\finpr}{\hfill $\square$ \vspace{2mm}}

\def\be{\begin{eqnarray}}

\def\ee{\end{eqnarray}}

\def\bee{\begin{eqnarray*}}

\def\eee{\end{eqnarray*}}

\newtheorem{thm}{Theorem}

\newtheorem{cor}{Corollary}

\newtheorem{lem}{Lemma}

\begin{document}

\title{ \vspace{-2cm} Efficient classical simulations of quantum Fourier transforms \\ and normalizer circuits over  Abelian groups}
\author{Maarten Van den  Nest\footnote{maarten.vandennest@mpq.mpg.de} \\ \\ {\normalsize Max-Planck-Institut f\"ur Quantenoptik,} \\ {\normalsize Hans-Kopfermann-Stra\ss e 1, D-85748 Garching, Germany. }}

\maketitle

\begin{abstract}
The quantum Fourier transform (QFT) is sometimes said to be the source of various exponential quantum speed-ups. In this paper we introduce a class of quantum circuits which cannot outperform classical computers even though the QFT constitutes an essential component. More precisely, we consider \emph{normalizer circuits}. A normalizer circuit over a finite Abelian group is any quantum  circuit comprising the QFT over the group, gates which compute automorphisms and gates which realize quadratic functions on the group. We prove that all normalizer circuits have polynomial-time classical simulations. The proof uses algorithms for linear diophantine equation solving and the monomial matrix formalism introduced in our earlier work. We subsequently discuss several aspects of normalizer circuits. First we show that our result generalizes the Gottesman-Knill theorem.
Furthermore we highlight connections to Shor's factoring algorithm and to the Abelian hidden subgroup
problem in general. Finally we prove that quantum factoring cannot be realized as a normalizer circuit owing to its modular exponentiation subroutine.
\end{abstract}

\section{ Introduction}

The quantum Fourier transform (QFT) is arguably the most important operation in the field of quantum algorithms. Indeed this transform is a principal component in several quintessential algorithms (cf. \cite{Jo98, Ch10} for  reviews), most importantly Shor's factoring algorithm \cite{Sh97}. In fact it is sometimes  said  that the QFT is the crucial non-classical ingredient accounting for the (believed) speed-up of quantum factoring.

Whereas in quantum algorithm design the focus lies on identifying scenarios where usage of the QFT yields new  quantum speed-ups, in this paper a complementary question is asked: under which conditions does a quantum algorithm, in spite of its use of the QFT, \emph{fail} to  outperform classical computation? Insight in this problem may shed light on the QFT's function in quantum computation  and indicate avenues for new quantum algorithms. In addition, since quantum algorithms have provided new ways to relate Fourier transforms to various problems in e.g. number theory and group theory (the central example being the factoring problem) a deeper understanding of the QFT might also lead to better \emph{classical} algorithms for these problems.

Whereas colloquially one often refers to ``the'' QFT as if it were a single operation,  a different QFT is in fact associated with every finite group. The standard QFT is defined over the Abelian group of integers modulo $2^n$. Throughout this work we focus on QFTs over arbitrary finite Abelian groups. This is arguably the class of QFTs which has been most successfully employed in quantum algorithms \cite{Ch10}.

We will introduce a class of quantum circuits, called here \emph{normalizer circuits}. A normalizer circuit over an Abelian group $G$ is any circuit containing the QFT over $G$ and two other types of operations, called automorphism gates and quadratic phase gates, both of which have natural group-theoretic definitions. In our main result we prove that all polynomial-size normalizer circuits, acting on a natural class of inputs (called coset states) and followed by a standard basis measurement, can be simulated classically in polynomial time.

This result sharply contrasts with the QFT's reputation of being classically intractable. In particular it highlights (complementary to previous works, cf. below) that the believed speed-up of e.g. Shor's algorithm cannot straightforwardly be attributed to the mere presence of the QFT: a more subtle mechanism must be responsible for this.  We will go into some detail analyzing the relation between normalizer circuits and quantum factoring. In particular we will prove that Shor's algorithm cannot be realized by a normalizer circuit, showing that (non-surprisingly) our result does not yield an efficient classical factoring algorithm. The single component of Shor's algorithm that is not covered by the normalizer gates formalism is  modular exponentiation, which is sometimes (mistakenly, in our opinion) considered to be a simple ``classical'' step.  In spite of this no-go result, we will highlight close connections between normalizer circuits and quantum factoring, and the Abelian hidden subgroup problem (HSP) in general\footnote{Recall that the HSP  encompasses the factoring problem as a special instance \cite{Bo95, Lo04}.}. Interestingly, the quantum states occurring in the HSP algorithm and those generated by normalizer circuits turn out to be very similar and, all in all, we find it very difficult to give a concise intuitive reason that clarifies why normalizer circuits have efficient classical simulations whereas the factoring algorithm (probably)  does not. In particular the usual superficial arguments that the power of quantum factoring comes from entanglement, interference or other ``quantumness'' measures, seem to apply equally well to normalizer circuits. Probably this indicates that such arguments have little merit and that much more delicate considerations are necessary.

Finally we remark that our results generalize the celebrated Gottesman-Knill theorem
which regards classical simulations of Clifford circuits [6]. In the language of the present
work, Clifford circuits correspond to normalizer circuits over the group $\mathbf{Z}_2\times ...\times \mathbf{Z}_2$.

\

\noindent {\bf Relation to previous work.} To our knowledge, classical simulations of QFTs were previously investigated in three independent works;  these results are similar to one another but different from ours.
In Ref. \cite{Ah06} it was shown that the approximate QFT over $\mathbf{Z}_{2^n}$ acting on a product input state and followed immediately by measurement in a product basis can be simulated classically in quasi-polynomial $O(n^{\log n})$ time. In Ref. \cite{Yo07} the same scenario was considered and a fully efficient classical simulation was achieved; more generally, efficient simulations were obtained for constant-depth circuits of bounded interaction range, interspersed with a constant number of approximate QFTs. In Ref. \cite{Br07} efficient simulations were obtained for the ``semi-classical'' QFT  acting on a class of entangled input states.  The common ingredient in these works
is their use of tensor contraction schemes which depend crucially on the \emph{structure} of the quantum circuit. These methods are only efficient when the graph representing the topology of the circuit is sufficiently close to a tree graph as measured by the tree-width \cite{Ma08}. A common feature of such circuits is that they can only generate limited amounts of entanglement (of certain type) \cite{Yo08}. The fundamental distinction with the present work is that our classical simulations are \emph{independent} of the structure of the circuit. This comes at the cost of restricting the allowed operations, similar to the Gottesman-Knill theorem.
Finally we mention that in \cite{Yo07b} it was shown that the existence of an efficient classical algorithm to simulate the modular exponentiation circuit acting on arbitrary product input states and for arbitrary product measurement, would imply an efficient simulation of Shor's algorithm. This result is similar in spirit to our result that modular exponentiation is the only component of quantum factoring that cannot be realized by a normalizer circuit.

\

\noindent {\bf Organization of this paper.}  Readers who are mainly interested in the results rather than their proofs may focus on sections \ref{sect_normalizer_circuits}-\ref{sect_relation_shor} where we introduce normalizer circuits, describe our main result and discuss its relation to Shor's algorithm and the HSP.   In subsequent sections mathematical tools are developed  and proofs of our main results are given. Considerable effort goes into laying out the necessary group theoretic concepts and their computational complexity, cf. section \ref{sect_preliminary}. In section \ref{sect_fund_Gcircuits} we discuss generalized Pauli and Clifford operators over Abelian groups. In section \ref{sect_monomial} we recall recently introduced techniques to describe and classically simulate a class of many body-states called M-states \cite{Va11}. In section \ref{sect_proof} we prove our main classical simulation result by combining the methods developed in sections \ref{sect_preliminary}-\ref{sect_monomial}. In section \ref{sect_shor} we prove that Shor's factoring algorithm cannot be realized as a normalizer circuit. Finally in section \ref{sect_further_examples} we give more examples of normalizer circuits.

\section{Normalizer circuits}\label{sect_normalizer_circuits}

Throughout this paper the symbol $G$ will represent a finite Abelian group given as a direct product of cyclic groups: \be\label{G} G=\mathbf{Z}_{d_1}\times\dots \times \mathbf{  Z}_{d_m}\ee where $\mathbf{Z}_{d_i}$ denotes the additive group of integers modulo $d_i$.  It is a fundamental result that every finite Abelian group is isomorphic to a decomposition of this form. Here we assume that $G$ is presented \emph{explicitly} in this way; this is a nontrivial constraint since computing this decomposition is generally a hard task. Every group element is a tuple ${  g} = g_1 \cdots g_m$ where $g_i\in \mathbf{  Z}_{d_i}$. Addition of  group elements is performed componentwise where arithmetic in the $i$-th component is modulo $d_i$.
The cardinality of $G$ is denoted by $\mathfrak{g}$.
For every $i$ we introduce a Hilbert space ${\cal H}_i$ with orthonormal basis $\{|x\rangle: x\in\mathbf{Z}_{d_i}\}$. Taking the tensor product of these $m$ spaces we obtain a $\mathfrak{g}$-dimensional space ${\mathcal H} $.
Defining $ |{  g}\rangle:= |g_1\rangle\otimes\dots\otimes|g_m\rangle$, the space ${\mathcal H}$ has a tensor product basis where each basis vector is labeled by an element of $G$. This basis will be called the standard basis. Note that the above construction of the space ${\cal H}$ is commonly used in the field of quantum algorithms (cf. e.g. \cite{Ch10}). We now define three classes of unitary operators on ${\cal H}$.

\vspace{2mm}

\noindent {\bf Quantum Fourier transforms.}
The quantum Fourier transform (QFT) over $\mathbf{  Z}_{d_i}$ is the following unitary operator on ${\cal H}_i$: \be\label{QFT_cyclic} F_i =\frac{1}{\sqrt{d_i}} \sum \omega_i^{xy} |x\rangle\langle y| ; \quad\quad  \omega_i = \exp(2\pi i/d_i) \ee where the sum is over all $x, y\in \mathbf{  Z}_{d_i}$. The QFT over the entire group $G$ is given by $F = F_1\otimes \cdots \otimes F_m$. Any operator obtained by replacing a subset of operators $F_i$ in this tensor product by identity operators will be called a partial QFT.

\vspace{2mm}

\noindent {\bf Automorphism gates.} An automorphism $\alpha$ is an isomorphism from $G$ to itself.  The associated automorphism gate $U_{\alpha}$ maps $|{  g}\rangle\to |\alpha({  g})\rangle$. Since every automorphism is invertible, $U_{\alpha}$ is a unitary operator which acts a permutation on the standard basis.

\vspace{2mm}

\noindent {\bf Quadratic phase gates.} A function $B$ from $G\times G$ to the nonzero complex numbers is said to be bilinear if for every $g, h, x \in G$ one has \be\label{bilinear} B(g+h, x) = B(g, x)B(h, x)\quad \mbox{and}\quad B(x, g+h) = B(x, g)B(x, h).\ee   A function $\xi$ from $G$ to the  nonzero complex numbers is said to be quadratic if there exists a bilinear function $B$ such that for every ${  g}, h\in G$ one has \be\label{quadratic_bilinear}\xi({  g}+ h) =\xi({  g})\xi(  h)B(g, h). \ee
Given a quadratic function $\xi$, the quadratic phase gate $D_{\xi}$ is the diagonal operator mapping $|{  g}\rangle\to \xi({  g})|{  g}\rangle$. In section \ref{sect_quadratic} we will show that $\xi({  g})$ is a complex phase for every ${  g}\in G$. This implies that every quadratic phase gate is a unitary operator.

\vspace{2mm}

A unitary operator which is either a (partial) quantum Fourier transform or its inverse, an automorphism gate or a quadratic phase gate will generally be referred to as a normalizer gate; the origin of this nomenclature will be clarified in section \ref{sect_fund_Gcircuits}. A quantum circuit composed entirely of normalizer gates is called a normalizer circuit over $G$. The size of a normalizer circuit is the number of normalizer gates of which it consists.

We further recall how computational cost is measured in the present context. A (classical or quantum) computation is called  efficient if the number of elementary operations of which it consists scales polynomially in the logarithm of $\mathfrak{g}$.  Since $\mathfrak{g}$ coincides with the dimension of ${\mathcal H}$, the above definition is compatible  with the standard concept of efficiency in quantum computation viz. the number of operations scales polynomially with the logarithm of the Hilbert space dimension.  In this work we will subsequently consider families of normalizer circuits of size polylog$(\mathfrak{g})$, henceforth called polynomial size normalizer circuits.

It is not evident that each individual normalizer gate can itself be  efficiently realized on a quantum computer i.e. as a quantum circuit composed of polylog$(\mathfrak{g})$ elementary gates.
It is well known that  (partial) quantum Fourier transforms over  arbitrary finite Abelian groups can be realized as efficient quantum circuits, either approximately \cite{Ki95} or exactly \cite{Mo04}. In appendix \ref{sect_app_efficient_circuits} we  show that all automorphism gates and quadratic phase gates can also be realized exactly as efficient quantum circuits.

\section{Examples}

We illustrate the definitions of normalizer gates with two  examples viz. $\mathbf{Z}_2^m$ and $\mathbf{Z}_{2^m}$. Both groups have  $\mathfrak{g}=2^m$ so the notion ``efficient'' is synonymous to ``in poly$(m)$ time''. See also section \ref{sect_further_examples} for further examples of normalizer gates.

\subsection{$\mathbf{Z}_2^m$ and Clifford circuits}\label{sect_examples_clifford}

The group $\mathbf{  Z}_2^m$ consists of all $m$-bit strings with addition modulo 2. Each Hilbert space ${\mathcal H}_i$ is a single qubit with basis $\{|0\rangle, |1\rangle\}$ so the total Hilbert space ${\cal H}$ is a system of $m$ qubits. The standard basis is the usual computational basis.  Applying definition (\ref{QFT_cyclic}) one finds that the QFT over $\mathbf{Z}_2$ is simply the Hadamard gate $H$. The QFT over $\mathbf{  Z}_2^m$ is  the $m$-fold tensor product of $H$ with itself and a partial QFT  acts as $H$ on some subset of the qubits and as the identity elsewhere. An automorphism of $\mathbf{  Z}_2^m$ is a linear map ${  x}\to A{  x}$ where $A$ is an invertible $m\times m$ matrix over $\mathbf{  Z}_2$. An example is the CNOT operation between any two bits.  An example of a quadratic function for $m=1$ is the function $x\to i^x$ where $x\in\mathbf{  Z}_2$. To verify that (\ref{quadratic_bilinear}) holds, one uses that  $i^{x + y} = i^x i^{y}(-1)^{xy}$  for every $x, y\in\mathbf{  Z}_2$, where $x+ y$ denotes the sum modulo 2.
For  $m=2$ the function $(x, y)\to (-1)^{xy}$ is quadratic as well, as can be easily verified. The  quadratic phase gates corresponding to the above functions are the single-qubit $\frac{\pi}{2}$-phase gate and the two-qubit controlled-Z gate, respectively: \be D= \mbox{ diag}(1, i);\quad  CZ= \mbox{ diag}(1, 1, 1, -1).\ee  For  arbitrary $m$, the gate $D$ acting on any single qubit and  CZ acting on any pair of qubits are quadratic phase gates as well. We can thus conclude that every quantum circuit composed of CNOT, $H$, CZ and $D$ gates is a normalizer circuit over $\mathbf{Z}_2^m$. We thus recover the well known class of  \emph{Clifford circuits}.

\subsection{$\mathbf{Z}_{2^m}$ and ``the'' QFT}\label{sect_examples_QFT}
Consider the group $\mathbf{  Z}_{2^n}$ consisting of integers modulo $2^n$.  The space ${\cal H}$  is  $2^n$-dimensional and the standard basis is labeled by the integers from 0 to $2^n-1$. The difference with $\mathbf{Z}_2^m$ is that the structure of $\mathbf{Z}_{2^m}$ does not naturally induce a factorization of ${\cal H}$ into $m$ single-qubit systems, since  the decomposition (\ref{G}) only contains a single cyclic group. As a consequence we will see that normalizer gates over $\mathbf{Z}_{2^m}$ will act \emph{globally} on ${\cal H}$, in contrast with $\mathbf{Z}_2^m$ where we found that certain normalizer gates are single- and two-qubit gates.

The QFT over $\mathbf{  Z}_{2^m}$ is the commonly known transform occurring in e.g. Shor's factoring algorithm. An explicit formula is obtained by replacing $d_i$ by $2^m$ in (\ref{QFT_cyclic}).  There are no partial QFTs since  the decomposition (\ref{G}) contains a single cyclic group.
Consider an arbitrary ${  a}\in \mathbf{  Z}_{2^m}$ which is coprime to $2^m$. Then
the multiplication map ${  x}\to {  ax}$ is an automorphism; the coprimality condition ensures that this map is invertible \cite{Ha80}. Hence $|x\rangle \to  |ax\rangle$ defines an automorphism gate. In section \ref{sect_further_examples} we show that  \be\begin{array}{c} D = \sum \ \omega^{{  b}{  x}^2 + cx} |{  x}\rangle\langle x| \quad\mbox{and}\quad  E = \sum\   \gamma^{  b  x(x + 2^m)} |{  x}\rangle\langle x|; \quad \omega = e^{2\pi i/2^m}, \ \gamma = \omega^{1/2}\end{array}\ee are  quadratic phase gates for arbitrary ${  b, c}\in \mathbf{  Z}_{2^m}$. In the definition of $E$, the quantity $b  x(x + 2^m)$ is computed over the integers.

\section{ Main result}\label{sect_main_result}

In our main result (theorem \ref{thm_main} below) we show that all polynomial size normalizer circuits can be efficiently simulated classically.  Before stating this result  we make some notions more precise.

Generally speaking, in the context of classical simulations the task is the following: given the description of a quantum circuit, its input and measurement, reproduce the output of the computation in polynomial time on a classical computer.  In the standard quantum circuit paradigm comprising local gates acting on at most $d$ qubits for some constant $d$, each gate can be naturally described by specifying on which $d$-qubit subsystem it acts and by listing all $2^d\times 2^d$ matrix elements; the description of the entire circuit is then essentially the concatenation of all descriptions of the gates in the circuit. In the context of normalizer circuits, however, the situation is different since these gates may act \emph{globally} on the entire $\mathfrak{g}$-dimensional Hilbert space (cf. the example of $\mathbf{Z}_{2^n}$ above). Therefore, specifying such operators in terms of their $\mathfrak{g}\times \mathfrak{g}$ matrix elements is in general exponentially inefficient and a different encoding should be used. In fact it is a priori not obvious that an efficient encoding, i.e.  comprising polylog$(\mathfrak{g})$ bits, exists at all for arbitrary normalizer gates.  Next we show that this is  the case.

First, a partial quantum Fourier transform  is simply described by the set of systems ${\cal H}_i$ on which it acts nontrivially.

Second we consider automorphism gates. For every $i$ ranging from 1 to $m$, denote by ${  e}^i\in G$ the group element which has $1\in\mathbf{Z}_{d_i}$ in its $i$-th component and zeroes elsewhere, where $0$ in slot $k$ represents the neutral element in $\mathbf{Z}_{d_k}$. Remark that ${  g} = \sum g_i {  e}^i $ for every ${  g}\in G$ so that the $m$ elements ${  e}^i$ generate $G$.
Therefore every automorphism $\alpha$ is fully specified by the $m\times m$ matrix  \be\label{matrix_endo} A:=[\alpha({  e}^1)|\cdots | \alpha({  e}^m)].\ee  We call $A$ the matrix representation of $\alpha$. Note that each column contains an element of $G$ and can thus be described by $O(\sum \log d_i) = O(\log \mathfrak{g})$ bits. Furthermore for every decomposition (\ref{G}) one has $m=O(\log \mathfrak{g})$  so that $A$ has $O(\log \mathfrak{g})$ columns \cite{Foot5}. Therefore,  $A$ yields an efficient description of  $U_{\alpha}$.

Third, letting $\xi$ be an arbitrary quadratic function, in section  \ref{sect_quadratic} we will show that

\vspace{2mm}

{\bf Q1.}  $\xi$ is completely determined by its action on the elements $e^i$ and  $e^i+e^j$;

{\bf Q2.} $\xi(g)^{2\mathfrak{g}}=1$ for every $g\in G$. Thus there exist $n(g)\in\mathbf{Z}_{2\mathfrak{g}}$ such that $\xi(g) = e^{\pi i n(g)/\mathfrak{g}}$.

\vspace{2mm}

\noindent It follows that the $O(m^2)$ integers $n(e^i)$ and $n(e^i+e^j)$ comprise an efficient description of $\xi$ and thus of the associated quadratic phase gate.

Henceforth we will assume that all normalizer gates are specified in terms of the descriptions given above, which will be called their \emph{standard encodings}. The standard encoding of a normalizer circuit consists of the concatenation of the standard encodings of the normalizer gates of which it is made up.

Finally, we will consider the following family of input states. If $K$ is a subgroup of $G$ and ${  x}\in G$, the associated coset state is  \be\label{coset_state} \begin{array}{c}|K+ {  x}\rangle = \frac{1}{\sqrt{K}} \sum |{  k}+{  x}\rangle,\end{array}\ee where the sum is over all $k\in K$. The standard encoding of this coset state is given by a generating set of $K$ of size polylog$(\mathfrak{g})$ (henceforth called a polynomial size generating set) together with the element ${  x}$. This description is efficient. Remark that in particular every standard basis state is a coset state.

\begin{thm}[{\bf Classical simulations of normalizer circuits}]\label{thm_main}
Let  $G=\mathbf{  Z}_{d_1}\times\dots \times \mathbf{  Z}_{d_m}$ be a finite Abelian group. Consider a polynomial size normalizer circuit over $G$ acting on a coset input state, where  both circuit and input are specified in terms of their standard encodings as described above. The circuit is followed by a measurement in the standard basis. Then there exists an efficient classical algorithm to sample the corresponding output distribution.
\end{thm}

Theorem \ref{thm_main} generalizes a well known result in the theory of classical simulations viz. the Gottesman-Knill theorem \cite{Go98}.  The latter regards classical simulations of Clifford circuits. The connection between normalizer circuits and Clifford circuits is obtained by considering the group $\mathbf{Z}_2^m$ as in  section \ref{sect_examples_clifford}.  Theorem \ref{thm_main} thus shows that every polynomial-size Clifford circuit acting on a coset input state and followed by a computational basis measurement, can be simulated efficiently classically. This recovers (a variant of) the Gottesman-Knill theorem.
Analogously, considering $\mathbf{  Z}_d^m$ where $d$ is any constant, theorem \ref{thm_main} recovers previously known generalizations of the Gottesman-Knill theorem from qubits to $d$-level systems \cite{Go99, Ho05}.

The novel content of theorem \ref{thm_main} regards all groups which \emph{cannot} be decomposed as a direct product $\mathbf{  Z}_d^m$ with $d$ constant. An interesting case is the single cyclic group $\mathbf{Z}_{2^m}$ as in section \ref{sect_examples_QFT}.  Theorem \ref{thm_main} implies that every  quantum circuit composed of poly$(m)$ QFTs over $\mathbf{Z}_{2^m}$, multiplication gates $|x\rangle\to |ax\rangle$ and diagonal gates $D$ and $E$, when acting on a coset input state and followed by standard basis measurement, can be simulated classically in poly$(m)$ time. More generally, one may also consider e.g. an $n$-fold direct product $\mathbf{Z}_{2^n}^m$  and construct the corresponding normalizer circuits. In this case ``efficient'' is synonymous to ``in poly$(n, m)$ time''. Theorem \ref{thm_main} implies that polynomial size normalizer circuits over this group can be efficiently simulated classically. We refer to section \ref{sect_further_examples} for more examples of normalizer gates.

Finally we point out two additional features of our results. First,  theorem \ref{thm_main} will yield an  efficient algorithm to sample the output distribution $\Pi$ of any normalizer circuit with \emph{perfect} accuracy. Second,  we will fully characterize $\Pi$  as follows: {\it There exists $x\in G$ and a subgroup $H\subseteq G$ such that $\Pi$  is the uniform distribution over the coset $H +x$}. Furthermore  $x$ and a  generating set of $H$ can be computed efficiently.

\section{Relation to Shor's factoring algorithm and the HSP}\label{sect_relation_shor}

We revisit Shor's factoring algorithm \cite{Sh97} in light of theorem \ref{thm_main}. Recall that this algorithm employs two quantum registers. Register 1 is a $2^m$-dimensional Hilbert space with standard basis states $|x\rangle$ with $x\in \mathbf{Z}_{2^m}$. Register 2 is $N$-dimensional with basis states $|y\rangle$ with $y\in \mathbf{Z}_N$ where $N$ is the number one wishes to factor. Fixing also $a\in\mathbf{Z}_N$ coprime to $N$, Shor's algorithm consists of the following steps \cite{Sh97, Ch10, Jo01}.

\vspace{2mm}

{\bf S1.} Initialize the total system in the state $|0, 0\rangle$.

{\bf S2.} Apply the QFT over $\mathbf{Z}_{2^m}$ to  register 1.

{\bf S3.} Apply the unitary operator $U_{\mbox{\scriptsize{me}}}$ defined by $|x, y\rangle  \to  |x, y+ a^x\rangle$.

{\bf S4.} Apply the QFT over $\mathbf{Z}_{2^m}$ to register 1.

{\bf S5.} Measure register 1 in the standard basis.

\vspace{2mm}

\noindent Repeating the above procedure $T$ times for some suitable $T$ scaling polynomially with the input size $\log N$ and suitably postprocessing the collected measurement outcomes with a classical computer yields a factor of $N$ in polynomial time, as proved by Shor. We now consider the class of normalizer circuits over $G=\mathbf{Z}_{2^m}\times \mathbf{Z}_N$ and ask whether S1-S5 are accounted for by theorem \ref{thm_main}. Except modular exponentiation S3, all steps are obviously covered. In section \ref{sect_shor} we prove that $U_{\mbox{\scriptsize{me}}}$ cannot be well-approximated by any normalizer circuit:
\begin{thm}\label{thm_shor_normalizer}
Let $N=pq$ where $p\geq 2$ and $q\geq 2$ are prime with $p\neq q$, let $a\in\mathbf{Z}_N$ be coprime to $N$ and consider $m\geq 2$. Consider the unitary operator $U_{\mbox{\scriptsize{me}}}$ defined in S3. There does not exist any (even exponential-size) normalizer circuit ${\cal C}$ satisfying $\|{\cal C} - U_{\mbox{\scriptsize{me}}}\|< 1-2^{-1/2}$ where $\|\cdot\|$ denotes the operator norm.
\end{thm}
Therefore theorem \ref{thm_main} does not lead to an efficient classical simulation of Shor's algorithm. But what is now ``the'' crucial element accounting for the speed-up of this algorithm? Is it the QFT---after all the only other component in the algorithm is modular exponentiation, which is a ``classical'' operation as it can entirely realized by e.g. a circuit of Toffoli gates? Or is it modular exponentiation---after all this is the only component not covered by the formalism of normalizer circuits? The point is that such questions do not have absolute answers as they are only meaningful in a relative setting: relative to the class of Toffoli circuits, the QFT is the essential ingredient that gives the factoring algorithm its power. By the same token, relative to the class of normalizer circuits, the essential ingredient is precisely modular exponentiation. The power of quantum factoring lies precisely it its ingenious combination of both elements.

We provide further connections between normalizer circuits and quantum factoring. Broadening our scope, we treat the hidden subgroup problem (HSP) for arbitrary Abelian groups \cite{Bo95, Ch10, Jo01}. In the HSP one has oracle access to a function $f:G\to \mathbf{Z}_N$ for some Abelian group $G$ and for some $N$, where it is promised that there exists an (unknown) subgroup $H\subseteq G$ such that $f(g)= f(h)$ iff  $g-h\in H$. The goal is to compute a generating set of $H$. Whereas the best classical algorithm requires $O(\mathfrak{g})$ operations, there exists a quantum algorithm which solves the problem in polylog$(\mathfrak{g})$ time. Very similar to Shor's algorithm, the HSP algorithm uses two  registers. Register 1 is  $\mathfrak{g}$-dimensional with standard basis states $|g\rangle$ where $g\in G$, and register 2 is $N$-dimensional with basis states $|y\rangle$ with $y\in \mathbf{Z}_N$. Letting $F$ denote the QFT over $G$, the steps in the HSP algorithm are fully analogous to S1-S5 as follows: initialize in $|0, 0\rangle$;  apply $F$ to register 1;  apply $U_f: |g, y\rangle\to |g, y+ f(g)\rangle$; apply $F$ to register 1; measure register 1.

Similar to Shor's algorithm, the quantum circuit for the HSP is not realized by any normalizer circuit because of $U_f$. However, let us consider the state $\rho$ of register 1 after application of $U_f$.  It is well known (and easily verified) that $\rho$ is the uniform mixture of all coset states of $H$: \be \rho =  \frac{1}{\mathfrak{g}}\sum_{x\in G} |H+x\rangle\langle H+x|.\ee Now remark that the only remaining operation in the HSP algorithm is the QFT, which is a normalizer gate. Revisiting theorem \ref{thm_main} we find that a computation with input $\rho$  followed by the application of the QFT can be simulated efficiently classically (by generating a random $x\in G$ and by classically simulating the QFT acting on the input $|H+x\rangle$). This appears to imply that the quantum algorithm for the HSP can be simulated efficiently classically by virtue of theorem \ref{thm_main}. However this conclusion is incorrect: the crucial point is that in the HSP  the state $\rho$ is \emph{unknown} as $H$ is unknown---indeed the entire purpose of the HSP  is to identify $H$! Therefore theorem \ref{thm_main} cannot be applied in this setting.

In short, the HSP can be embedded in the \emph{same} computational model considered in theorem \ref{thm_main}, which has normalizer circuits acting on (mixtures of) coset state(s) as its main ingredients. The crucial distinction  is that in the HSP the input state $\rho$ is unknown and the goal is to identify it, as in a tomography problem. This viewpoint also highlights the important role of the \emph{first} stage of the HSP algorithm whose purpose is the preparation of the unknown $\rho$ via application of $U_f$ -- even though this first stage appears to be a mere ``classical'' step! Another implication is that we cannot expect to learn much about the speed-up achieved by the HSP algorithm (and thus in particular Shor's algorithm) by quantifying, say, how much entanglement is generated during the subroutine $\rho\to F\rho F^{\dagger}$. Indeed the exact same process with a fully known but otherwise identical input state $\rho$ \emph{can} be simulated classically.

\section{Preliminaries on Abelian groups}\label{sect_preliminary}

Here we discuss concepts in group theory such as characters, endomorphisms and quadratic functions, as well as the computational complexity of some group theoretic tasks.

\subsection{Characters and orthogonal groups}\label{sect_char}
We review some basic representation theory of finite Abelian groups; see \cite{Fu91} for more details. The set ${\cal F}$ of all complex functions $f:G\to \mathbf{C}$ is well known to be a $\mathfrak{g}$-dimensional Hilbert space where the inner product between two functions $f_1$ and $f_2$ is defined by $\langle f_1, f_2\rangle := \sum \bar f_1(g)f_2(g)$; here the sum is over all $g\in G$ and  $\bar f_1(g)$ is the complex conjugate of $f_1(g)$.   A character of  $G$ is a function $\chi\in {\cal F}$  satisfying $\chi(g)\neq 0$ for every $g$ and\be \chi({  g}+ h)=\chi({  g})\chi( h)\quad \mbox{ for every } {  g}, h\in G.\ee  Equivalently, a character is a homomorphism from $G$ to the multiplicative group of nonzero complex numbers.  The trivial character maps $g\to 1$ for every $g$.
For every $g=g_1\cdots g_m$, define the following complex function on $G$:
\begin{eqnarray}\label{chi}
\begin{array}{c}
\chi_g: h=h_1\cdots h_m\   \to \   \exp 2\pi i\left( \frac{g_1h_1}{d_1}+\dots +\frac{g_mh_m}{d_m} \right).
\end{array}
\end{eqnarray}

\begin{lem}[{\bf Characters}]\label{thm_chi}
For all $g, h\in G$ and for all characters $\chi$ and $\psi$ the following holds:

\vspace{2mm}

(a) If $\chi\neq \psi$ then $\langle \chi, \psi\rangle = 0$ i.e. these characters are orthogonal.

(b) $\chi_g$ is a character of $G$.

(c) $\chi_g(h) = \chi_h(g)$.

(d) $\chi_{g+h} = \chi_g\chi_h$.

(e) Define $S(g):=\sum_{h\in G}\chi_g(h)$. Then $S(0)=\mathfrak{g}$ and $S(g)=0$ for every $g\neq 0$.

(f) Every character is a quadratic function.

(g) The QFT over $G$ satisfies $F = \frac{1}{\sqrt{\mathfrak{g}}}\sum \chi_g(h)|g\rangle\langle h|$, where the sum is over all $g, h\in G$.

(h) Every character of $G$ is one of the functions $\chi_{{  g}}$ i.e. $G$ has precisely $\mathfrak{g}$ characters.
\end{lem}
{\it Proof:} since $\chi$ is a homomorphism one has $\chi(0) = 1$. Since $\mathfrak{g}g=0$ for every $g\in G$ and since $\chi$ is a character it follows that $1 = \chi(0)=\chi(\mathfrak{g}g) = \chi(g)^{\mathfrak{g}}$. This shows that $\chi(g)$ is a complex phase for every $g\in G$. If $\chi$ is different from $\psi$ there exists $g_0\in G$ such that $\chi(g_0)\neq \psi(g_0)$ and thus, since $\chi(g_0)$ is a complex phase, $\bar \chi(g_0)\psi(g_0)\neq 1$. Then \be \langle \chi, \psi\rangle &=& \sum \bar\chi(g)\psi(g) = \sum \bar\chi(g+g_0)\psi(g+g_0) \nonumber\\ &=& \bar\chi(g_0)\psi(g_0)  \sum  \bar\chi(g)\psi(g) =  \bar\chi(g_0)\psi(g_0)  \langle \chi, \psi\rangle \ee where sums are over all $g\in G$, and claim (a) follows. Claims (b)-(c)-(d) follow straightforwardly from definition (\ref{chi}).  To prove (e) note that $S(g)$ coincides with the inner product between the character $\chi_g$ and the trivial character. Furthermore $\chi_g$ is the trivial character iff $g=0$. Using (a) then yields (e).
To prove (f) remark that (\ref{quadratic_bilinear}) is satisfied by taking $B$ to be the trivial bilinear function which sends all $(g, h)$ to 1. Claim (g) is obtained straightforwardly by using the definitions of $\chi_g$ and $F$.   Let ${\cal F}$ denote the $\mathfrak{g}$-dimensional  space of all complex functions on $G$ as above. Since every two characters are orthogonal owing to (a), there can be at most $\mathfrak{g}$ characters. Owing to (b) all functions $\chi_g$ are characters; furthermore $\chi_g=\chi_h$ iff $g=h$ so that there are precisely $\mathfrak{g}$ distinct functions $\chi_g$. It follows that there can be no other characters; this proves (h).
\finpr

For every subgroup $H\subseteq G$ define the following subset of $G$: \be H^{\perp} := \{{  g}\in G: \chi_{{  g}}({  h})=1 \mbox{ for every } {  h} \in H\}.\ee

\begin{lem}[{\bf Orthogonal groups}]\label{H_perp}
For every subgroup $H\subseteq G$ the following holds.

\vspace{2mm}

(a) $H^{\perp}$ is a subgroup of $G$, called the orthogonal group of $H$.

(b) $|H^{\perp}|\cdot |H| =\mathfrak{g}$.

(c) $(H^{\perp})^{\perp} = H$.
\end{lem}
{\it Proof:} (a) is proved straightforwardly by using that each $\chi_g$ is a character. To prove (b),  for every $g\in G$ let $\psi_g:H\to \mathbf{C}$ be the restriction of $\chi_g$ to $H$ i.e. $\psi_g(h) = \chi_g(h)$ for every $h\in H$. The function $\psi_g$ is easily seen to be a character of $H$. Define \be \tau(g) := \sum_{h\in H} \chi_g(h) = \sum_{h\in H}\psi_g(h).\ee Remark that $\tau(g)$ is the inner product between the trivial character of $H$ and the character $\psi_g$. Lemma \ref{thm_chi}a implies that $\tau(g)=|H|$ if $\psi_g$ is the trivial character on $H$ and $\tau(g)=0$ otherwise. Furthermore the definition of $H^{\perp}$ implies that $\psi_g$ is the trivial character precisely when $g\in H^{\perp}$. Thus $\tau(g)=|H|$ iff $g\in H^{\perp}$. Now consider the state $|H\rangle$ as in (\ref{coset_state}). Then (recalling lemma \ref{thm_chi}g) \be\label{fourier_H} F|H\rangle = \frac{1}{\sqrt{g|H|}}  \sum_{g\in G}\sum_{h\in H} \chi_g(h) |g\rangle= \frac{1}{\sqrt{g|H|}}\sum_{g\in G} \tau(g)|g\rangle =  \frac{\sqrt{|H|}}{\sqrt{g}} \sum_{g\in H^{\perp}} |g\rangle.\ee Since $F$ is unitary, the norm of $F|H\rangle$
 is 1. Together with (\ref{fourier_H}) this yields (b).
Finally we prove (c). By definition one has $g\in H^{\perp\perp}$ iff $\chi_g(k)=1$ for all $k\in H^{\perp}$. Consider $h\in H$. Then by definition of $H^{\perp}$ one has $\chi_k(h)=1$ for all $k\in H^{\perp}$. Using lemma \ref{thm_chi}c it follows that $h\in H^{\perp\perp}$ and we conclude that $H\subseteq H^{\perp\perp}$. Applying claim (b)  to $H$ and $H^{\perp}$ yields $|H||H^{\perp}| = \mathfrak{g} = |H^{\perp}||H^{\perp\perp}|$ so that $|H| = |H^{\perp\perp}|$. But then we must have  $H = H^{\perp\perp}$. \hfill $\square$

\subsection{Endomorphisms}

An endomorphism $\varpi$ of $G$ is a homomorphism from $G$ to itself i.e. $\varpi(g+h)=\varpi(g) + \varpi(h)$ for all $g, h\in G$. Every endomorphism $\varpi$ has an $m\times m$ matrix representation similar to (\ref{matrix_endo}). We characterize which matrices can occur a matrix representations of endomorphisms.

\begin{lem}[{\bf Matrix representation}]\label{thm_matrix_endo}
Let $A$ be an $m\times m$ matrix where the entries in the $k$-th row of $A$ belong to $\mathbf{Z}_{d_k}$ for every $k$. Let $a^i\in G$ denote the $i$-th column of $A$. Then there exists an endomorphism of $G$ with  matrix representation $A$  if and only if $d_ia^i=0$ for every $i$.
\end{lem}
{\it Proof:} let  $A$ be the matrix representation of an endomorphism $\varpi$. Since $d_i {  e}^i= {  0}$, it follows that $0 = \varpi(0) = \varpi(d_i e^i)= d_i \varpi({  e}^i)$ and so $d_i a^i=0$. We prove the converse. Consider a matrix $A$ satisfying $d_ia^i=0$ for every $i$. Define $\varpi({  g}):=\sum_i g_i {  a}^i$ for every ${  g}=g_1\cdots g_m\in G$. We prove that $\varpi $  is an endomorphism. Consider arbitrary ${  g}, h\in G$.  We introduce the following notation: the sum of $g_i$ and $h_i$ over $\mathbf{  Z}_{d_i}$ is denoted by $g_i \oplus h_i$. The sum of these two numbers regarded as \emph{integers} is denoted by $g_i+h_i$. With this notation,  applying the definition of $\varpi$ one obtains \be \label{endo1} \varpi({  g}+ {  h}) &=& \sum (g_i\oplus h_i) {  a}^i.\\ \label{endo2} \varpi({  g})+ \varpi({  h}) &=& \sum g_j {  a}^i  + \sum h_i {  a}^i = \sum (g_i+ h_i) {  a}^i.\ee Since the $i$-th component of ${  g}+ {  h}$ is $g_i \oplus h_i = g_i + h_i \mbox{ mod } d_i$, there exists an integer $q_i$ such that $g_i \oplus h_i = g_i + h_i + q_id_i$. Since $d_i {  a}^i = {  0}$, it follows that $(g_i+ h_i) {  a}^i =  (g_i\oplus h_i) {  a}^i$.
Combining this last equation with (\ref{endo1}) and (\ref{endo2}) shows that $\varpi$ is an endomorphism. \finpr

Consider an endomorphism $\varpi$, fix ${  g}\in G$ arbitrarily and consider the map $\varphi_g = \chi_g \cdot \varpi $: \be\label{map_transpose}\varphi_g: {  x\in G} \to \chi_{{  g}}(\varpi({  x})).\ee Using that $\chi_{{  g}}$ is a character and that $\varpi$ is an endomorphism one shows that $\varphi_g$ is also a character. Recalling lemma \ref{thm_chi}h there must hence exists a unique  $\hat{g}$ such that $\varphi_g = \chi_{\hat g}$. Let $\hat{\varpi}$ denote the map which sends ${  g}$ to $\hat{  g}$. Using lemma \ref{thm_chi}d it is straightforward to show that $\hat\varpi$ is also an endomorphism. Indeed for every $x\in G$ one has \be \varphi_{g+h}(x) = \chi_{g+h}(\varpi(x)) = \chi_g(\varpi(x))\chi_h(\varpi(x)) = \varphi_g(x)\varphi_h(x)\ee implying that $\varphi_{g+h}= \varphi_g\varphi_h$ and thus $\hat\varpi(g+h)=\hat\varpi(g) + \hat\varpi(h)$ for every $g, h\in G$. The map $\hat\varpi$ is henceforth called here the dual endomorphism of $\varpi$.

\begin{lem}[{\bf Dual matrix representation}]\label{thm_matrix_dual}
Let $A$ be the matrix representation of an endomorphism $\varpi$. Then the matrix representation $B$ of $\hat\varpi$ is given by \be \frac{d_k}{d_l} A_{lk} \mbox{ mod } d_k = B_{kl} \quad \mbox{for every } k, l=1, \dots, m.\ee
\end{lem}
{\it Proof:}  we have $\chi_{{  e}^l}(\varpi({  e}^k))= \chi_{\hat\varpi({e}^l)}({  e}^k)$ for every  $k, l$ by definition of $\hat\varpi$. Using that $A$  and $B$ are the matrix representations of $\varpi$ and $\hat \varpi$, respectively, this yields \be \begin{array}{c}\exp \frac{ 2\pi i A_{lk}}{d_l} = \exp \frac{2\pi i B_{kl}}{d_k}.\end{array}\ee  It follows that $B_{kl}$ is equal to $d_k A_{lk}/d_l$ up to addition of an integer multiple of $d_k$. This implies that
\be d_k A_{lk}/d_l \mbox{ mod } d_k = B_{kl} \mbox{ mod } d_k.\ee
Since $B$ is the matrix representation of an endomorphism, $B_{kl}$ is an integer between 0 and $d_{k}-1$. This implies that $B_{kl}$  mod $d_k = $ $B_{kl}$ and the result follows.
\finpr

To illustrate the above result, consider a group $G$ of the form $\mathbf{Z}_d^m$ i.e. all cyclic groups in the decomposition of $G$ have the same order. Then lemma \ref{thm_matrix_dual} implies that the matrix representation of the dual endomorphism is simply the transposed matrix $A^T$. Finally, we remark that lemma \ref{thm_matrix_dual} can be used to show that $(\hat{\varpi})\hat{}=\varpi$. The argument is straightforward.

\subsection{Quadratic functions}\label{sect_quadratic}
We prove  claims Q1-Q2 made in section \ref{sect_main_result} by means of the following lemma.

\begin{lem}[{\bf Quadratic and bilinear functions}]\label{thm_quadratic_phase}
(a) A function $B$ from $G\times G$ to the nonzero complex numbers is bilinear if and only if there exists an endomorphism $\varpi$ of $G$ such that $B(g, h) = \chi_{\varpi(g)}(h)$. Furthermore for every bilinear function $B$ and quadratic function $\xi$ as in (\ref{quadratic_bilinear}), for every $g,h\in G$ and for every positive integer $n$, the following holds:
\vspace{2mm}

(b) $B(g, 0)=1=B(0, g)$.

(c) $B(g, h)^{\mathfrak{g}}=1$.

(d) $\xi(0)=1$.

(e) $\xi(n{  g}) = \xi({  g})^n \  B(g, g)^{f(n)}\quad \mbox{ where } f(n) = n(n-1)/2$.

(f) $\xi({  g})^{2\mathfrak{g}}=1$.

\end{lem}
{\it Proof:} consider an endomorphism $\varpi$. Using lemma \ref{thm_chi}b,d and the fact that $\varpi$ is an endomorphism, one shows that $B(g, h) = \chi_{\varpi(g)}(h)$ is bilinear. Conversely, consider an arbitrary bilinear function $B$. By definition of bilinearity,  for every fixed $g$ the function $\beta_g: h\to B(g, h)$ is a character. Lemma \ref{thm_chi}h implies that there exists a unique $g^*$ such that $\beta_g = \chi_{g^*}$. Let $\varpi$ denote the map which sends $g$ to $g^*$. Using that $g\to B(g, h)$ is also a character and  lemma \ref{thm_chi}d it follows that $\varpi$  must be an endomorphism.  This proves (a). Claim (b)  follows from (a) and the fact that $\varpi(0)=0$ for every endomorphism. Claim (c) is proved by using (a) and by noting that $\chi_x(y)^{\mathfrak{g}}=1$ for every $x, y\in G$ where the latter follows directly from definition (\ref{chi}). Using that $B(0, 0)=1$ one finds \be \xi(0)= \xi(0+0)=\xi(0)\xi(0)B(0, 0) = \xi(0)^2.\ee Recalling in addition that by definition $\xi(g)\neq 0$ for every $g$ yields (d).
Claim (e) is proved by induction on $n$; the statement is obviously correct for $n=0 ,1$. Assuming (e) holds for some integer $n$, one has  \be\xi((n + 1)g) = \xi(ng) \xi(g) B(ng, g) =  \xi(ng) \xi(g) B(g, g)^n = \xi({  g})^{n+1} \  B(g, g)^{f(n)+n} \ee In the first identity we used that $\xi$ is quadratic, in the second that $B$ is bilinear; in the final identity we used the induction hypothesis. Noting that $f(n)+ n = f(n+1)$ yields the proof.
Using (e) with $n=2\mathfrak{g}$ and recalling (c) yields  $\xi(2\mathfrak{g}{g}) = \xi({  g})^{2\mathfrak{g}}$. Noting that $2\mathfrak{g}{  g}={  0}$ and recalling (d) proves (f).  \finpr

Lemma \ref{thm_quadratic_phase}f proves that $\xi({  g})$ is a complex phase.  Repeatedly applying the definitions of quadratic and bilinear functions yields, for every $g=\sum g_i e^i\in G$:
\be
\xi(g)  &=&  \prod_i \xi(g_ie^i) \prod_{i<j} B(e^i, e^j)^{g_ig_j}\label{xi_g_1}\\
B(e^i, e^j) &=& \xi(e^i+e^j)\ \overline{\xi(e^i)}\ \overline{\xi(e^j)}. \label{xi_g_2}
\ee
Furthermore,  lemma \ref{thm_quadratic_phase}e implies
\be
\begin{array}{c}
\xi(g_ie^i)  = \xi(e^i)^{g_i} B(e^i, e^i)^{f(g_i)}.\label{xi_g_3}
\end{array}
\ee
We conclude  that $\xi(g)$ is fully determined by the values $\xi(e^i)$ and $\xi(e^i+e^j)$.

If $\xi$ is a quadratic function where the associated $B$ is given as in lemma \ref{thm_quadratic_phase}a, we will simply say that $\xi$ is a quadratic function with endomorphism $\varpi$. For every quadratic function, the associated $B$ and  $\varpi$ are unique. Uniqueness of $B$ follows from (\ref{quadratic_bilinear}). To show uniqueness of $\varpi$, assume that $B(g, h) = \chi_{\varphi(g)}(h)$ for some other endomorphism $\varphi$. Then \be 1 = B(0, h) = B(g-g, h)= B(g, h)B(-g, h) = \chi_{\varpi(g)}(h)\chi_{-\varphi(g)}(h) = \chi_{\varpi(g) - \varphi(g)}(h),\ee for all $g, h\in G$. This implies that $\varpi=\varphi$.

\subsection{Computations in Abelian groups} We show that several problems involving the concepts introduced earlier in this section have efficient classical algorithms. In the following the \emph{size} of an integer is the number of bits in its binary expansion. Recall that for every group  (\ref{G}) one has $m=O(\log \mathfrak{g})$. Remark also that  the size of each $d_i$ is $O(\log \mathfrak{g})$ since $d_i\leq \mathfrak{g}$ and thus for every $g=g_1\cdots g_m\in G$ the size of each $g_i$ is O$(\log \mathfrak{g})$ as well.

We start by discussing some elementary tasks.  Given two integers $a$ and $b$ of size at most $l$, their sum, product as well as the remainder  $a$ mod $b$ can be computed in poly$(l)$ time. Hence the sum $g+h$ can be computed in polylog$(\mathfrak{g})$ time for every $g, h\in G$ by computing the $m$ remainders $g_i + h_i $ mod $d_i$. Similarly, $ng$ can be computed in polylog$(\mathfrak{g}, n)$ time for every integer $n$  by computing the remainders $ng_i$ mod $d_i$.

Let $\varpi$ be an endomorphism given in terms of its matrix representation. Remarking that $\varpi(g)=\sum g_i \varpi(e^i)$ for every $g=g_1\cdots g_m$, the group element $\varpi(g)$ can be computed efficiently on input of $g$. Lemma \ref{thm_matrix_dual} shows that the matrix representation of $\hat\varpi$ can be computed efficiently as well.

Let $\xi$ be a quadratic function given in terms of its standard encoding. Consider the associated $B$ as in (\ref{quadratic_bilinear}) and  $\varpi$ as in lemma \ref{thm_quadratic_phase}a. Then the matrix representation $A$ of $\varpi$ can be computed efficiently: one first computes the quantities $B(e^k, e^l)$ using (\ref{xi_g_2}). Using \be B(e^k, e^l) = \chi_{\varpi(e^k)}(e^l) = e^{2\pi i A_{lk}/d_l}\ee yields the matrix element $A_{lk}$.  Furthermore the function value $\xi(g)$ can be computed efficiently  on input of $g$: first compute all quantities $B(e^i, e^j)$ using (\ref{xi_g_2}), then compute $\xi(g_i e^i)$ using (\ref{xi_g_3}), finally compute $\xi(g)$ using (\ref{xi_g_1}).

Next we discuss more involved tasks:.

\begin{thm}[{\bf Efficient computations in Abelian groups}]\label{thm_inverse_orthogonal}
Consider a finite Abelian group (\ref{G}). Then the following problems can be solved in polylog$(\mathfrak{g})$ time:

\vspace{2mm}

(a) Given a generating set of a subgroup $H$, compute a generating set of $H^{\perp}$.

(b) Given  $h^1, \dots, h^r\in G$ with $r=$ polylog$(g)$ and $s_1, \dots, s_r\in\mathbf{Z}_{\mathfrak{g}}$, determine whether

\quad \ \ there exists $g\in G$ satisfying
\be \chi_{h^k}(g) = e^{2\pi i s_k/\mathfrak{g}} \quad \mbox{ for all } k\ee

\quad  \ \ and, if yes, compute such a solution $g$.

(c) Given the matrix representation of an automorphism $\alpha$, compute the matrix

\quad \ \ representation of $\alpha^{-1}$.
\end{thm}
\noindent Property (a) is known in the quantum algorithms community, cf. the Abelian HSP algorithm \cite{Lo04}. Property (b) is similar to (a). It is however not clear to the author if (c) is known.  A proof of theorem \ref{thm_inverse_orthogonal} is given in appendix \ref{sect_app_inverse_orthogonal}. For now we mention that a central ingredient in the proof is the fact that systems of linear diophantine equations can be solved efficiently, as recalled next.

\begin{thm}[{\bf Linear Diophantine equations} \cite{Bl66, Ga76, Ka79, Ch82}]\label{thm_diophantine}
Let $A$ be an $n\times m$ matrix and let $b$ be an  $n$-dimensional vector, both with integer entries of size at most $k$. Both $A$ and $b$ are specified by listing all of their entries in binary. Then there exists an algorithm with running time poly$(n, m, k)$ which decides whether the equation $Ax=b$ has an integer solution $x$. Furthermore, if a solution exist, then there exist integer vectors $x^0$ and $x^1, \dots, x^r$, for some $r\leq m$, such that:

\vspace{2mm}

(a) $Ax^0=b$.

(b) $Ax^i=0$ for all $i=1, \dots, r$.

(c) Every integer solution to $Ax=b$ has the form $x= x^0+y$ where $y$ is an integer

\quad \ \ linear combination of the $x^i$.

\vspace{2mm}

\noindent Furthermore all components of $x^0$ and the $x^i$ have size poly$(n, m, k)$ and can be computed in poly$(n, m, k)$ time.
\end{thm}

\section{Pauli and Clifford operators over Abelian groups}\label{sect_fund_Gcircuits}

Here we consider the Hilbert space ${\cal H}$ associated with $G$ as in section \ref{sect_normalizer_circuits} and define several families of operators acting on this space.

For every  ${  g}\in G$  define the  pair of operators \be\label{X_g} X({  g}) = \sum |h+g\rangle\langle h|; \quad Z({  g}) = \sum \chi_{  g}({  h}) |{  h}\rangle\langle h|\ee where the sums are over all $h\in G$. Remark that $X(g)$ is a permutation matrix and that $Z(g)$ is a diagonal unitary operator.
Let $a\in\mathbf{Z}_{2\mathfrak{g}}$ and consider the following $2\mathfrak{g}$-th root of unity $\gamma:= e^{i\pi /\mathfrak{g}}$. An operator of the form $\sigma = \gamma^a Z({  g})X({  h})$ will be called a  Pauli operator over $G$, or Pauli operator in short, as this definition  generalizes the standard notion of Pauli matrices defined on qubit systems \cite{Foot6}. The triple $(a, {  g}, {  h})$ is called the label of $\sigma$.
It is important that the label constitutes a description of $\sigma$ comprising $O(\log \mathfrak{g})$ bits i.e. this description is efficient.
The following commutation relations  are verified straightforwardly by applying (\ref{X_g}):
\be\label{commutation}
X({  g})X({  h}) &=& X({  g}+{  h}) = X({  h})X({  g}),\nonumber\\ Z({  g})Z({  h})&=&Z({  g}+{  h})=Z({  h})Z({  g}),\nonumber\\ Z({  g}) X({  h}) &=& \chi_{  g}({  h})X({  h})Z({  g}).\quad \ee
The first equation in (\ref{commutation}) implies that the set $\{X({  g}):{  g}\in G\}$ is a group w.r.t. the standard operator product. In fact, this group is isomorphic to $G$: the isomorphism is the map ${  g}\to X({  g})$.  Similarly, $\{Z({  g}):{  g}\in G\}$ is a group isomorphic to $G$.

Elementary manipulations with Pauli operators are computationally efficient in the following sense:

\begin{lem}[{\bf Products and powers of Pauli operators}]\label{thm_pauli_products_powers}
Consider Pauli operators $\sigma$ and $\tau$ and a positive integer $n$. Then  $\sigma\tau$, $\sigma^n$ and $\sigma^{\dagger}$ are also Pauli operators, the labels of which can be computed in polylog$(\mathfrak{g}, n)$ time on input of $n$ and the labels of $\sigma$ and $\tau$. Moreover,  $\sigma^{\dagger}= \sigma^{2\mathfrak{g}-1}$.
\end{lem}

{\it Proof:} let $(a, {  g}, {  h})$ and $(b, x, y)$ denote the labels of $\sigma$ and $\tau$, respectively.  Using the commutation relations (\ref{commutation}) one finds  \be \label{sigma_product}\sigma\tau = \gamma^{a+b} \cdot \chi_{  h}(  x) \cdot X(g+x)Z(h+ y).\ee The overall phase on the r.h.s. of (\ref{sigma_product}) has the form $\gamma^s$ for some $s\in\mathbf{Z}_{2\mathfrak{g}}$, so that $\sigma \tau$ is indeed a Pauli operator. It is straightforward that $s$ as well as $g+x$ and $h+y$ can be computed efficiently. The other claims in the lemma are proved analogously. \finpr

Lemma \ref{thm_pauli_products_powers} implies that the set of all Pauli operators forms a finite group, called here the Pauli group (over $G$). A unitary operator $U$ on ${\cal H}$ is called a Clifford operator (over $G$)   if $U$ maps the Pauli group onto itself under the conjugation map $\sigma\to U\sigma U^{\dagger}$. It is easy to see that the set of all Clifford operators forms a group. Formally speaking, the Clifford group is the normalizer of the Pauli group in the full unitary group acting on ${\cal H}$.
We now show that normalizer gates and Clifford operators are fundamentally interrelated.
\begin{thm}[{\bf Normalizer gates are Clifford}]\label{thm_G_circuit_fundamental}
Every normalizer gate is a Clifford operator. Furthermore let $U$ be a normalizer gate specified in terms of its standard classical encoding as in section \ref{sect_normalizer_circuits} and let $\sigma$ be a Pauli operator specified in terms of its label. Then the label of $U \sigma U^{\dagger}$ can be computed in polylog($\mathfrak{g}$) time.
\end{thm}
{\it Proof:} we show that $UX({  g}) U^{\dagger}$ and $UZ({  g}) U^{\dagger}$ are Pauli operators with efficiently computable labels for every ${  g}$. The proof of the theorem then follows from the fact that every Pauli operator is given as a product $\gamma^{a} X({  g})Z({  h})$ and by using lemma \ref{thm_pauli_products_powers}.

Let $F$ be the QFT over $G$, consider an automorphism $\alpha$ and a quadratic function $\xi$ with endomorphism $\varpi$. We describe the action of $F$, $U_{\alpha}$  and $D_{\xi} $ on  $X({  g})$ and $Z({  g})$ under conjugation; $\alpha^{-*}$ is shorthand for $(\alpha^{-1})^{*}$:
\be\label{normalizer_Clifford} \begin{array}{ccll}
F&: & X({  g})\to Z({  g}); & Z({  g})\to X({  -g})\vspace{1mm}\\
U_{\alpha}&: & X({  g})\to X(\alpha({  g})); & Z({  g})\to Z(\alpha^{-*}({  g}))\vspace{1mm}\\
D_{\xi}& : & X({  g})\to \xi({  g}) X({  g}) Z(\varpi({  g})) ; \ \  & Z({  g})\to Z({  g})
\end{array} \ee

We prove that these are the correct actions. One has
\be
FX(g)F^{\dagger}|h\rangle &=& \frac{1}{\mathfrak{g}} \sum_{k, l} \overline{\chi_k(h)}\chi_{k+g}(l)|l\rangle = \frac{1}{\mathfrak{g}}\sum_l \chi_g(l) \sum_{k} \chi_{k}(l-h)|l\rangle \label{fourier2}\\
&=& \frac{1}{\mathfrak{g}}\sum_l \chi_g(l) \sum_{k} \chi_{l-h}(k)|l\rangle = \chi_g(h)|h\rangle = Z(g)|h\rangle.\label{fourier4}\ee The first identity is obtained using lemma \ref{thm_chi}g; using  $\bar \chi_k(h) = \chi_k(-h)$ and lemma \ref{thm_chi}b, d yields the second identity;  the third identity uses lemma \ref{thm_chi}c; the fourth is proved using lemma \ref{thm_chi}e.  The action of the Fourier transform on $Z(g)$ is computed analogously and the argument is omitted.

Using that $U_{\alpha}^{\dagger} = U_{\alpha^{-1}}$ yields \be U_{\alpha}X(g)U_{\alpha}^{\dagger}|h\rangle &=& |\alpha( \alpha^{-1}(h) + g)\rangle = | h + \alpha(g)\rangle=  X(\alpha(g)) |h\rangle.\ee The action on $Z(g)$ is computed analogously by applying the definition  of the dual automorphism. Finally, one has \be U_{\xi} X(g) U_{\xi}^{\dagger} |h\rangle &=& \xi(g+h)\overline{\xi(h)} |g+h\rangle= \xi(g) B(g, h)|g+h\rangle \\ &=&\xi(g) \chi_{\varpi(g)}(h)|g+h\rangle =\xi(g)X(g)Z(\varpi(g))|h\rangle,\ee where we in the second identity we have used (\ref{quadratic_bilinear}) and in the third lemma \ref{thm_quadratic_phase}a.

We now show that the actions of normalizer gates on Pauli operators can be computed efficiently. The argument heavily relies on theorem \ref{thm_inverse_orthogonal} as well as the discussion above it.
The action of $F$ can obviously be computed efficiently. As for $U_{\alpha}$, recall that the action $g\to \alpha(g)$ can be computed efficiently. As the matrix representations of inverse and dual automorphisms can be computed efficiently, the action $g\to \alpha^{-*}(g)$ can be computed efficiently as well. As for $D_{\xi}$,  recall that both $\xi(g)$ and $\varpi(g)$ can be computed efficiently on input of $g$ together with the standard representation of $\xi$.

Finally, analogous to the arguments above, it can be shown that the action of partial QFTs can be computed efficiently as well; this argument is omitted. \finpr

Variants of the above result were previously known for groups of the form $\mathbf{Z}_{d}^m$ with $d$ constant (see \cite{Go99, Ho05} and references within), although those works do not use the terminology of normalizer gates, automorphism gates, quadratic phase gates, etc.

We conclude by remarking that every Pauli operator can be realized as a polynomial size (in fact: constant size) normalizer circuit, up to a global phase. To see this, first note that  $Z(g)$ is a quadratic phase gate since the functions $\chi_g$ are quadratic functions (recall lemma \ref{thm_chi}f). As for $X(g)$, recall from (\ref{normalizer_Clifford}) that $X(g)  = F^{\dagger} Z(g) F$. But then every Pauli operator is realized as a constant-size normalizer circuit as well.

\section{M-states}\label{sect_monomial}

Next we recall results of \cite{Va11} where a formalism was introduced to describe and classically simulate a class of quantum states called M-states. The brief account given here is self-contained and will suffice for our purposes.

As before we let $G$ be an Abelian group with associated Hilbert space ${\cal H}$. A unitary operator $U$ on ${\cal H}$ is \emph{monomial} if $U=PD$ for some  diagonal matrix  $D$ and some  permutation matrix $P$, both relative to the standard basis.  A state $|\psi\rangle$ is an \emph{M-state} if there exists a group ${\cal G} $ consisting entirely of monomial unitary matrices such that $|\psi\rangle$ is the unique eigenvector with eigenvalue 1 of all elements of ${\cal G}$ simultaneously.  The group  ${\cal G}$  is said to be a monomial stabilizer group of $|\psi\rangle$. In theorem \ref{thm_M_state1} below we characterize the structure of an M-state $|\psi\rangle$ in terms of the properties of ${\cal G}$. To do so we first develop  some further concepts.
Consider $U\in G$ and $g\in G$; then, since $U$ is monomial,  there exists $h\in G$ such that $U|g\rangle \propto |h\rangle$. The group thus ${\cal G}$ naturally acts on the elements of $G$ via the map $g\to h$. The  orbit of $g$ under this action is the set of all $h$ that can be reached by applying all possible elements from ${\cal G}$: \be\label{orbit} O_g = \{h: \exists U\in{\cal G}\mbox{ s.t. } U|g\rangle\propto|h\rangle\}.\ee
Finally, for every $g\in G$ let ${\cal G}_{g}$ be the subset of ${\cal G}$ consisting of all $U\in {\cal G}$ satisfying $U|g\rangle\propto |g\rangle$ i.e. $U$ acts trivially on $g$; this subset is easily seen to be a subgroup of ${\cal G}$.
\begin{thm}[{\bf M-states} \cite{Va11}]\label{thm_M_state1}
Let $|\psi\rangle$ be an M-state with monomial stabilizer group ${\cal G}$. Then the following holds:

\vspace{2mm}

(a) $|\psi\rangle$ is a uniform superposition. This means that all non-zero  amplitudes $\langle g|\psi\rangle$ are

\quad \ \ equal in modulus.

(b) Let $S\subseteq G$ be the support of $|\psi\rangle$ i.e. $S$ is the collection of all $g$ with $\langle g|\psi\rangle\neq 0$.

\quad \ \ Then $S$ coincides with one of the orbits $O_x$.

(c) Consider an arbitrary $x\in G$ and let $V_1, \dots,V_r$ be generators of ${\cal G}_x$. Then one has \be S = O_x \quad \mbox{iff} \quad V_{i}|x\rangle = |x\rangle \mbox{ for all } i.\ee
\end{thm}

\noindent For completeness, theorem \ref{thm_M_state1} is proved  in appendix \ref{sect_app_M_states}. The relevance of this result for the proof of theorem \ref{thm_main} is the following. We will show that the output state $|\psi\rangle$ of any normalizer circuit is always an M-state. Owing to theorem \ref{thm_M_state1}a-b, a standard basis measurement on $|\psi\rangle$ yields a uniformly random element of  $O_x$. We will thus prove theorem \ref{thm_main} by  showing that the orbit $O_x$ can be determined efficiently (using theorem \ref{thm_M_state1}c) and, subsequently, that a random element in $O_x$ can be generated classically efficiently as well.

\section{Proof of theorem \ref{thm_main}}\label{sect_proof}

Consider a polynomial size normalizer circuit ${\cal C}$ acting on the input $|K+x\rangle$. Note that $|K+x\rangle = X(x)|K\rangle$. Since $X(x)$ can be realized as a constant-size normalizer circuit (recall section \ref{sect_fund_Gcircuits}), this operator may be absorbed into  ${\cal C}$. Thus henceforth in this section we consider inputs of the form $|K\rangle$, without loss of generality.

\subsection{The input state}

We start by characterizing the input state as a joint eigenvector of a family of Pauli operators. We say that a state $|\psi\rangle$ is a ``+1 common eigenvector'' of a set ${\cal S}$ of operators if the state is an eigenvector with eigenvalue 1 for every operator in the set; $|\psi\rangle$ is the ``unique +1 common eigenvector'' of ${\cal S}$ if it is, up to a global phase, the only vector with this feature.

\begin{lem}\label{thm_input}
The set ${\cal G}_0$ consisting of all operators $Z({  u})X({  v})$ with ${  u}\in K^{\perp}$ and ${  v}\in K$ is an Abelian group. Furthermore $|K\rangle$ is the unique +1 common eigenvector of ${\cal G}_0$.
\end{lem}
{\it Proof: } using the definitions of Pauli operators and their commutation relations it is easily verified that ${\cal G}_0$ is an Abelian group and that $|K\rangle$ is a +1 common eigenvector of ${\cal G}_0$. To show uniqueness, let $|\psi\rangle$ be an arbitrary +1 common eigenvector of ${\cal G}_0$, so that \be \langle g|Z(u)X(v)|\psi\rangle=\langle g|\psi\rangle \quad\mbox{ for every } {  u}\in K^{\perp},   v\in K, g\in G.\ee  Using (\ref{X_g}) this implies that \be \label{K_common_eigenstate} \chi_{u}(g) \langle g+v|\psi\rangle = \langle g|\psi\rangle.\ee
For every $g\notin K$ there exists $u\in K^{\perp}$ such that $\chi_{{  u}}({  g})\neq 1$: indeed, if this were not the case then $g$ would belong to $K^{\perp\perp}$; recalling that $K^{\perp\perp}= K$ (see lemma \ref{H_perp}c) then contradicts with the assumption that $g\notin K$. It then follows from (\ref{K_common_eigenstate}) with $v=0$ that $ \langle {  g}|\psi\rangle = 0$ for every $g\notin K$. Further, taking $u=0=g$ and $v\in K$ arbitrary in (\ref{K_common_eigenstate}) implies that $\langle {  v}|\psi\rangle=\langle {  0}|\psi\rangle$. This shows that $|\psi\rangle$ must be proportional to $|K\rangle$.  \hfill $\square$

\subsection{The output state is an M-state}

Lemma \ref{thm_input} implies that the final state ${\cal C}|K\rangle =:|\psi\rangle$ is the unique +1 common eigenvector of the group ${\cal G}$  obtained by applying the conjugation map $A\to {\cal C}A{\cal C}^{\dagger}$ to all elements of ${\cal G}_0$. Theorem \ref{thm_G_circuit_fundamental} shows that every element of ${\cal G}$ is a Pauli operator.
Remark that every Pauli operator $\sigma$ is monomial and unitary as it is given as the product of a diagonal unitary matrix $\gamma^a Z(g)$ and a permutation matrix (and hence unitary) $X(h)$.  Thus $|\psi\rangle$ is an M-state with monomial stabilizer group ${\cal G}$.

Theorem \ref{thm_M_state1} implies that $|\psi\rangle$ is a uniform superposition and that the support of $|\psi\rangle$ is one of the orbits $O_x$ as in (\ref{orbit}). Next we characterize these orbits. If $\sigma$ is a Pauli operator with label $(a, {  g}, {  h})$, we call $g$ the ``$Z$-component'' and $h$ the ``$X$-component'' of $\sigma$. Denote the $X$-component formally by $\varphi(\sigma):= {  h}$.  Now let $H\subseteq G$ be the image of ${\cal G}$ under the map $\varphi$ i.e. $H$ is the set of all $X$-components of the elements of ${\cal G}$.  The commutation relations (\ref{commutation}) readily  yield \be\label{X_component} \varphi(\sigma\tau)= \varphi(\sigma) + \varphi(\tau)\quad\mbox{ for all } \sigma, \tau \in {\cal G}.\ee
This implies that $H$ is a subgroup of $G$.
Using that $\sigma|x\rangle \propto |x+h\rangle$ for every $\sigma\in {\cal G}$ with label $(a, g, h)$, one straightforwardly shows that
\be O_x = x+H\ee for every $x\in G$. Thus these orbits are precisely the cosets of the subgroup $H$.
Theorem \ref{thm_M_state1}a-b now immediately implies:

\begin{cor}\label{thm_uniform_sup}
There exists $x^0\in G$ s.t. $|\psi\rangle$ is a uniform superposition with support $H+ x^0$.
\end{cor}

\noindent Let $\Pi$ denote the probability distribution resulting from measuring $|\psi\rangle$ in the standard basis. We have arrived at a crucial insight for the classical simulation we aim to achieve:
\begin{cor}\label{thm_pi}
$\Pi$ is the uniform distribution over the coset $H+{  x}^0$.\end{cor}

\subsection{Computing a generating set of $H$}

Next we show that a generating set of $H$ can be computed efficiently. Let $\{{  u}^1, \dots, {  u}^k\}$  generate $K$ with $k=O(\log \mathfrak{g})$. This set is assumed to be specified as an input.  Owing to theorem \ref{thm_inverse_orthogonal}a, a generating set $\{{  u}^{k+1}, \dots, {  u}^n\}$ of $K^{\perp}$ can be computed efficiently, for some $n=O(\log \mathfrak{g})$. It follows that ${\cal G}_0$ is generated by the operators \be X({  u}^1), \dots, X(u^k), Z({  u}^{k+1}), \dots, Z(u^n).\ee Consequently, a generating set of ${\cal G}$ is obtained by conjugating these operators with ${\cal C}$: \be \sigma_i  := \left\{\begin{array}{ll} {\mathcal C} X({  u}^i) {\mathcal C}^{\dagger}& \quad \mbox{ for every }i = 1, \dots, k\\ {\mathcal C} Z({  u}^i) {\mathcal C}^{\dagger}& \quad \mbox{ for every }i= k+1, \dots, n.\end{array}\right.\ee  The $\sigma_i$ mutually commute since they are obtained by simultaneously conjugating a set of mutually commuting operators. Theorem \ref{thm_G_circuit_fundamental} shows that every $\sigma_i$ is a Pauli operator with efficiently computable label, say $(a_i, {  g}^i, {  h}^i)$. Remark that (\ref{X_component}) implies that $\varphi: {\cal G}\to G$ is a group homomorphism.  Since ${\cal G}$ is generated by the $\sigma_i$ and since $\varphi$ is a homomorphism, the image of $\varphi$ is generated by the elements $\varphi(\sigma_i) = h^i$. But this image is precisely the subgroup $H$.

\subsection{Computing $x^0$}

Next we show that a suitable $x^0$ as in corollary \ref{thm_uniform_sup} can be computed efficiently. Our approach will be to use theorem \ref{thm_M_state1}c. To this end we first characterize the subgroups ${\cal G}_x\subseteq {\cal G}$ which will turn out to have a particularly simple form; in particular we will show that all these groups are \emph{equal}. Define the set ${\cal D}$ to consist of all $\sigma\in{\cal G}$ that, up to some global phase, have the form $Z({  g})$ for some ${  g}\in G$. It is easily verified that this set is a subgroup of ${\cal G}$, called the diagonal subgroup.
\begin{lem} \label{thm_D_support}
Every group ${\cal G}_x$ is equal to the diagonal subgroup ${\cal D}$.
\end{lem}
{\it Proof:} since every $D\in {\cal D}$ is a diagonal operator, one has $D|x\rangle\propto|x\rangle$, showing that ${\cal D}\subseteq {\cal G}_x$. Conversely, if $\sigma\in{\cal G}_x$  has label $(a, g, h)$ then $\sigma |x\rangle\propto |x+h\rangle$. Since $\sigma\in {\cal G}_x$ the state $|x\rangle$ is an eigenvector of $\sigma$; this can only be true if $h=0$, showing that $\sigma\in {\cal D}$.  \hfill $\square$

\begin{lem}\label{thm_D_generating_set}
The labels of a generating set of ${\cal D}$ can be computed efficiently.
\end{lem}
{\it Proof:} since the $\sigma_i$ mutually commute,  the map  \be T: k=k_1\cdots k_n\in \mathbf{Z}^n \ \to \  \sigma_1^{k_1}\dots \sigma_n^{k_n} \in {\cal G}\ee
satisfies $T(k+k')=T(k)T(k')$. This shows that $T$ is a group homomorphism from the additive group $\mathbf{Z}^n$ to ${\cal G}$. Consider the inverse image of ${\cal D}$ under $T$: \be D = \{k\in \mathbf{Z}^n: T(k)\in {\cal D}\}.\ee Since $T$ is a homomorphism, $D$ is a subgroup of $\mathbf{Z}^n$. Furthermore we have $T(D)={\cal D}$: the inclusion $\subseteq$ holds by definition of $D$; to show $\supseteq$, note that the $\sigma_i$ generate ${\cal G}$ so that every element in ${\cal G}$ can be written as $T(k)$ for some $k$; therefore every element in ${\cal D}$ can be written as $T(k)$ with $k\in D$. Our strategy to obtain a generating set of ${\cal D}$ will be to first compute a generating set of $D$. Since $T(D)={\cal D}$, the images of these generators under $T$ will generate ${\cal D}$.

Applying (\ref{commutation}) one finds
\be\label{T_k} T(k)\propto Z\left(\sum k_i{  g}^i\right)X\left(\sum k_i{  h}^i\right).\ee It follows that $k\in D$ if and only if  $\sum k_i{  h}^i = {  0}.$ This last identity is satisfied iff
\be\label{condition_D}  k_1 h^1_{j}  + \dots + k_n h^n_{j}  \equiv 0 \mbox{ mod } d_j\quad\mbox{ for every } j=1, \dots, m \quad (\mbox{where } {  h}^i = h^i_{1}\dots h^i_{m}).\ee This in turn holds iff  there exists an integer vector $l = l_1\cdots l_m$ such that
\be\label{dioph}  k_1 h^1_{j}  + \dots + k_n h^n_{j}  + d_jl_j = 0\quad\mbox{ for every } j.\ee
Remark that (\ref{dioph}) is a system of $m$ linear diophantine equations in the $n+m$ unknowns $(k, l)$. Furthermore $n = O(\log \mathfrak{g}) = m$ and  $h_j^i$ and $d_j$ are integers of size  $O(\log \mathfrak{g})$.  Using theorem \ref{thm_diophantine}, a generating set of  solutions \be (k^{\alpha}, l^{\alpha}), \quad \alpha= 1, \dots, r \quad \mbox{for some } r = \mbox{ polylog}(\mathfrak{g})\ee can thus be computed in polylog($\mathfrak{g}$) time. But then every solution $k$ to (\ref{condition_D}) is generated by the elements $k^{\alpha}$, implying that the latter form a generating set of $D$.

As argued above, it now follows that ${\cal D}$ is generated by the operators $T(k^{\alpha})$. Finally, given $k^{\alpha}$ the label of $T(k^{\alpha})$ can be computed efficiently by repeatedly applying lemma \ref{thm_pauli_products_powers}, since $T(k^{\alpha})$ is defined as a product of powers of the Pauli matrices $\sigma_{i}$.
\finpr

We use shorthand notation $T(k^{\alpha})=: V_{\alpha}$ for the generators  of ${\cal D}$ computed in lemma \ref{thm_D_generating_set}.   Since $V_{\alpha}$ belongs to the diagonal subgroup, its label is $(c_{\alpha}, z^{\alpha}, 0)$ for some  $c_{\alpha}\in\mathbf{Z}_{2\mathfrak{g}}$ and ${  z}^{\alpha}\in G$, all of which are efficiently computable owing to lemma \ref{thm_D_generating_set}.

In summary so far, we have computed a generating set of ${\cal D}$ and thus (via lemma \ref{thm_D_support}) of every subroup ${\cal G}_x$. Now let $S$ denote the support of $|\psi\rangle$ and consider an arbitrary $x\in G$. Theorem \ref{thm_M_state1}c then implies that $S= x+H$ if and only if $V_{\alpha}|x\rangle = |x\rangle$ for every $\alpha$. This last condition holds iff \be\label{linear_eq} \chi_{{  z}^{\alpha}}({  x}) = \gamma^{-c_{\alpha}}\quad\mbox{ for every } \alpha=1, \dots, r\ee where we have used that \be V_{\alpha}|x\rangle = \gamma^{c_{\alpha}}Z(z^{\alpha})|x\rangle = \gamma^{c_{\alpha}} \chi_{{  z}^{\alpha}}({  x})|x\rangle.\ee
The  equations (\ref{linear_eq}) are guaranteed to be satisfied by \emph{some} $x$ since we know from corollary \ref{thm_uniform_sup} that the support $S$ coincides with some coset $x+H$.  Since the l.h.s. of (\ref{linear_eq}) are $\mathfrak{g}$-th roots of unity, these equations can only have a solution if $\gamma^{-c_{\alpha}}$ is also a $\mathfrak{g}$-th root of unity. Thus there must exist $s_{\alpha}\in\mathbf{Z}_{\mathfrak{g}}$ such that $\gamma^{-c_{\alpha}} = e^{2\pi i s_{\alpha}/\mathfrak{g}}$. Given $c_{\alpha}$, the number $s_{\alpha}$ can easily be computed. In conclusion,  we find that \be S = x+H \quad \mbox{ iff }\quad \chi_{{  z}^{\alpha}}({  x}) = e^{2\pi i s_{\alpha}/\mathfrak{g}}\quad\mbox{ for every } \alpha=1, \dots, r.\ee Theorem  \ref{thm_inverse_orthogonal}b can now be used to compute a solution ${  x}\equiv x^0$ efficiently.

\subsection{Sampling from $H+ x^0$}

Having determined the coset $H+x^0$ coinciding with the support of $|\psi\rangle$, we show that a random element in this coset can be generated efficiently. Recalling corollary \ref{thm_pi} this will complete the proof of theorem \ref{thm_main}.
\begin{lem}
A random element of $H+x^0$ is generated in polylog$(\mathfrak{g})$ time as follows:

\vspace{2mm}

(a) Generate  $t_i\in\mathbf{Z}_{\mathfrak{g}}$ uniformly at random, where $i=1, \dots, n$.

(b) Compute $h:= \sum t_i h^i$

(c) Output $h+x^0$.
\end{lem}
{\it Proof: }  first we show that the procedure (a)-(b) generates a uniformly random element of $H$. Remark that this procedure  can only output elements within $H$ and, since $H$ is generated by the $h^i$,  each element of this subgroup occurs with nonzero probability. We show that all these probabilities are equal; some care is required since  the generators $h^i$ are not guaranteed to be independent. The map $\nu: \mathbf{Z}_{\mathfrak{g}}^n\to H$ defined by $\nu(t):=\sum t_ih^i$  (where $t = t_1\dots t_n$) is easily seen to be a homomorphism between the additive group $\mathbf{Z}_{\mathfrak{g}}^n$ and  $H$. With (a)-(b) every $h\in H$ is generated with probability  \be p(h) = |\{t :\nu(t)=h\}|/\mathfrak{g}^n.\ee
The kernel $N$ of $\nu$ is a subgroup of $\mathbf{Z}_{\mathfrak{g}}^n$ since $\nu$ is a homomorphism. Let $t_h\in \mathbf{Z}_{\mathfrak{g}}^n$ be a single vector satisfying $\nu(t_h)=h$ and consider the coset $t_h +N$. We now claim that \be\label{sets_equality}t_h + N = \{t: \nu(t)=h\}.\ee The inclusion $\subseteq$ holds trivially. To prove the converse inclusion, note that every $t$ satisfying $\nu(t)=h$ can be written as $t=t_h + t-t_h$, where $t-t_h\in N$. This shows (\ref{sets_equality}).  Note further that each coset of $N$ has the same cardinality as $N$. Therefore $p(h)$ is equal to $| N|/|\mathfrak{g}^n|$ and thus independent of $h$. This shows that this probability distribution is uniform over $H$. It easily follows that (a)-(b)-(c) generates a uniformly random element in $H + x^0$. Finally, it is easily verified that this procedure can be implemented efficiently.  \hfill $\square$

\section{Proof of theorem \ref{thm_shor_normalizer}}\label{sect_shor}

Suppose that there exists a normalizer circuit ${\cal C}$ satisfying $\|{\cal C} - U_{\mbox{\scriptsize{me}}}\|<\delta$ with $\delta= 1-1/\sqrt{2}$. Our goal is to arrive at a contradiction.

Defining  $F(x, y) :=(x, y+ a^x)$, the inequality $\|{\cal C} - U_{\mbox{\scriptsize{me}}}\|<\delta$ implies that $\| {\cal C}|x, y\rangle - |F(x, y)\rangle\| < \delta$ and thus \be\label{mod_exp} |\langle F(x, y)|{\cal C}|x, y\rangle|>1-\delta = 1/\sqrt{2}\ee for every $x$ and $y$. Since ${\cal C}|x, y\rangle$ is a coset state, owing corollary \ref{thm_uniform_sup} this state must be a uniform superposition. If this state is a uniform superposition of two or more standard basis states, then $|\langle u, v|{\cal C}|x, y\rangle|\leq 1/\sqrt{2}$ for all standard basis states $|u, v\rangle$, yielding a contradiction with  (\ref{mod_exp}). We therefore conclude that ${\cal C}|x, y\rangle$ is a just a single basis state. In particular (\ref{mod_exp}) implies that ${\cal C}|x, y\rangle$ must then coincide with $|F(x, y)\rangle$ for every $x$ and $y$ and thus ${\cal C} = U_{\mbox{\scriptsize{me}}}$. We now use the following lemma, to be proved at the end of this section:
\begin{lem}\label{thm_mod_exp}
Consider an arbitrary finite Abelian group $G$. Consider a unitary gate $U: |g\rangle\to |F(g)\rangle$, where $F$ is a permutation of $G$, which can be realized as a normalizer circuit. Then there exists an automorphism $\alpha$ and  $t\in G$ such that $F(g)=\alpha(g)+ t$ and $F$ is called \emph{affine}.
\end{lem}
\noindent Since we have showed that ${\cal C} = U_{\mbox{\scriptsize{me}}}$, lemma \ref{thm_mod_exp} implies that the exponential function $F(x, y) =(x, y+ a^x)$ must be an affine function on the group $G= \mathbf{Z}_{2^m}\times \mathbf{Z}_N$. We show that this is not the case, implying that our initial assumption about the existence of ${\cal C}$ was false. Suppose that there exists an automorphism $\alpha$ of $G$ with matrix representation $A$ and $t\in G$ such that $F(x, y) = \alpha(x, y) + t$. By evaluating this identity in $(0, 0)$, $(1, 0)$ and $(0, 1)$ is easy to show that then $t=(0, 1)$ and \be A = \left[ \begin{array}{cc} 1& 0\\a-1& 1\end{array}\right].\ee The identity $F(x, y) = \alpha(x, y) + t$ then implies  that \be\label{mod_exp_proof} a^x \equiv (a-1)x+1 \mbox{ mod } N \quad\mbox{ for all } x\in\{0, \dots, 2^m-1\}. \ee Considering $x=2$ yields $(a-1)^2 \equiv 0$ mod $N$. Thus, since $a\neq 1$, there exists a nonzero integer $k$ such that $(a-1)^2=kN$. Since $N=pq$ with $p\neq q$ it follows that both $p$ and $q$ divide $(a-1)^2$. But then $p$ and $q$ also divide $a-1$.  In other words,  both $p$ and $q$ occur in the prime decomposition of $a-1$. It follows that $N=pq$ divides $a-1$. This leads to a contradiction since $a\leq N$. This proves theorem \ref{thm_shor_normalizer}.

\

We conclude by proving lemma \ref{thm_mod_exp}. Suppose that $U$ can be realized as a normalizer circuit. Then for every $g\in G$ there exists $\alpha(g)\in G$ such that $UX(g)U^{\dagger} = X(\alpha(g))$.  Indeed since $U$ acts as a permutation on the standard basis and since $X(g)$ is also a permutation, $UX(g)U^{\dagger}$ is again a permutation; since by assumption $U$ maps $X(g)$ onto a Pauli operator and since only the Pauli operators $X(h)$ are permutations,  the claim follows. Using that \be UX(g)U^{\dagger}\  UX(g')U^{\dagger} = UX(g+g')U^{\dagger}\ee one shows that $\alpha$ is an automorphism. Applying both sides of the identity $UX(g)U^{\dagger} = X(\alpha(g))$ to $|0\rangle$ yields \be\label{shor_2} F(g + F^{-1}(0)) = \alpha(g) \mbox{ for every } g\ee Setting $h:= g + F^{-1}(0)$ and $t:= - \alpha(F^{-1}(0))$ and using that $\alpha$ is an automorphism yields $F(h) = \alpha(h) +t$ for every $h\in G$. This proves lemma \ref{thm_mod_exp}.

\section{Further examples of normalizer gates}\label{sect_further_examples}

In this final section we give further illustrations of group automorphisms and quadratic functions, and the associated normalizer gates. See also section \ref{sect_normalizer_circuits}.

We first consider automorphisms. Consider an arbitrary $a\in\mathbf{Z}_{d}$. Then the multiplication map $M_a: x\to ax$ is an endomorphism of the cyclic group $\mathbf{Z}_{d}$. If  $a$ is coprime to $d$ then $a$ has a multiplicative inverse i.e. there exists $b$ in $\mathbf{Z}_{d}$ such that $ab=1$ \cite{Ha80}.  Therefore, for such $a$ the multiplication map is an automorphism. Second, consider the group $\mathbf{Z}_{d_1}\times \mathbf{Z}_{d_2}$, fix $c\in\mathbf{Z}_{d_2} $ satisfying $d_1c\equiv 0$ mod $d_2$ and define \cite{Foot9}:\be\label{auto_C} F_{c}: (x, y)\in \mathbf{Z}_{d_1}\times \mathbf{Z}_{d_2}  \ \to\ (x, y+xc).\ee Owing to lemma \ref{thm_matrix_endo}, the map $F_c$ is an endomorphism. Furthermore this map has an inverse viz. $F_{-c}$, showing that $F_c$ is an automorphism. Remark that for $d_1=d_2=2$ and $c=1$ one obtains the CNOT gate. Considering an arbitrary Abelian group  (\ref{G}), applying $M_a$  to any cyclic group $\mathbf{Z}_{d_i}$ (with $a$ coprime to $d_i$) and applying  $F_c$  to any pair $\mathbf{Z}_{d_i}\times \mathbf{Z}_{d_j}$ (with $d_ic\equiv 0$ mod $d_j$) yield automorphisms of $G$. Finally, since the set of all automorphisms of any group is a group as well, further examples  can be obtained by taking products and inverses (and combinations thereof) of the examples given above.

Next we give examples of quadratic functions. Consider $a\in \mathbf{Z}_d$. One can straightforwardly verify that the functions \be\label{ex_quadratic_1} x\in\mathbf{Z}_d &\to& e^{ 2 \pi iax/d} \quad\mbox{ and }\quad x\in\mathbf{Z}_d \ \to \  e^{2 \pi i ax^2/d}\label{ex_quadratic_1'}\ee are quadratic functions for $\mathbf{Z}_d$. Remark that (\ref{ex_quadratic_1}) coincides with the character $\chi_a$. Second, consider $\mathbf{Z}_{d_1}\times \mathbf{Z}_{d_2}$ and $c\in\mathbf{Z}_{d_2}$ satisfying $d_1c\equiv 0$ mod $d_2$. Then \be\label{ex_quadratic_2} (x, y)\in \mathbf{Z}_{d_1}\times \mathbf{Z}_{d_2} \ \to \  e^{2\pi i cxy/d_2}\ee is quadratic. This can be shown by using the following lemma:
\begin{lem}\label{thm_quadratic_prop} Consider an arbitrary finite Abelian group. For every endomorphism $\varpi$, the  following function is quadratic: $f_{\varpi}(g):= \chi_g(\varpi(g))$.
\end{lem}
{\it Proof: } the proof is obtained straightforwardly by using basic properties of $\chi_g$ and the fact that $\varpi$ is an endomorphism.
\finpr

Using lemma \ref{thm_matrix_endo} one notes that $\varpi: (x, y)\to (0, cx)$ is an endomorphism. The associated function $f_{\varpi}$ is precisely (\ref{ex_quadratic_2}), showing that the latter is indeed quadratic.  Remark that for $d_1=d_2=2$ and $c=1$ the function  (\ref{ex_quadratic_2}) maps the pair of bits $(x,y)$ to $(-1)^{xy}$. The associated  quadratic phase gate is the CZ gate, cf. section \ref{sect_normalizer_circuits}. Lemma \ref{thm_quadratic_prop} allows to generate an entire class of quadratic functions. Interestingly, not every quadratic function is of the form $f_{\varpi}$. To see this, consider again a single cyclic group and denote $q_a(x):= ax( x + d)$ where $a, x\in\mathbf{Z}_d$ but $q_a(x)$ is computed is \emph{over the integers} (i.e. not within $\mathbf{Z}_d$). Then  \be\label{ex_quadratic_3} x\ \to\ e^{\pi i q_a(x)/d}\ee is a quadratic as well; the proof is straightforward but somewhat tedious and we refer to appendix \ref{sect_app_quadratic}. For $d=2$ and $a=1$ one simply has $x\to (-i)^x$ which gives rise to the single-qubit phase gate diag$(1, -i)$ as discussed in section \ref{sect_normalizer_circuits}. Note that the function (\ref{ex_quadratic_3}) cannot be of the form $f_{\varpi}$. Indeed the function values of the former are  $2\mathfrak{g}$-th roots of unity, whereas the function values of the latter are $\mathfrak{g}$-th roots of unity.

Applying (\ref{ex_quadratic_1}) and (\ref{ex_quadratic_3}) to any single cyclic subgroup of a general Abelian group $G$ and and applying  (\ref{ex_quadratic_2}) to any pair of cyclic subgroups yield quadratic functions of $G$.  Finally we remark that the set of a quadratic functions of $G$ can be shown to be a group under the multiplication of functions; the neutral element is the trivial function $g\to 1$ and the inverse of $\xi$ is its complex conjugate. Thus further examples of quadratic functions (and the associated quadratic phase gates) can be obtained by taking products and complex conjugates, and combinations thereof, of the examples given above.

\subsection*{Acknowledgements}
I thank J. Bermejo-Vega for discussions, W. D\"ur for suggestions to improve the presentation of the manuscript, and A. Schrijver, A. Sch\"onhage and A. Storjohann for providing references on linear diophantine equations.

\appendix

\section{Efficient quantum circuits}\label{sect_app_efficient_circuits}

We show that every automorphism gate and quadratic phase gate can be realized as an efficient quantum circuit.

Let $\alpha$ be an automorphism of $G$, specified in terms of its matrix representation.  Recall that the function $\alpha$  is invertible and that $\alpha(g)$ is efficiently computable on input of $g$ (cf. section \ref{sect_preliminary}). Thus there exists an efficient circuit of classical reversible gates that computes $g\to \alpha(g)$. But then the standard translation of these classical gates into unitary operators yield an efficient quantum circuit to compute $|g\rangle\to |\alpha(g)\rangle$.

Let $\xi$ be a quadratic function, specified in terms of its standard representation as in section \ref{sect_normalizer_circuits}. Recall from section \ref{sect_preliminary} that $g\to \xi(g)$ can be computed efficiently classically. Furthermore each $\xi(g)$ is a $2\mathfrak{g}$-th root of unity. Writing $\xi(g) = \exp[ i \pi  n(g)/\mathfrak{g}]$ where $n(g)$ is an integer between 0 and $2\mathfrak{g}$, we denote by $f(g)$ the bit string corresponding to the binary expansion of $n(g)$. Note that each $f(g)$ has length $k=O(\log \mathfrak{g})$ and that $g\to f(g)$ can be computed efficiently classically. We will write $f(g)=f_1(g)\cdots f_k(g)$ where each $f_i(g)$ is a bit; the relation to $n(g)$ is given by $n(g) = \sum_{j=1}^k f_j(g)2^{j-1}$. It follows that \be\label{xi_f(g)} \xi(g) = \prod \omega_j^{f_j(g)}; \quad \omega:= \exp[{ i \pi  2^{j-1}/\mathfrak{g}}].\ee
Now consider the following procedure: start with the input $|g\rangle |0\rangle$ where $|0\rangle$ represents a string of $k$ ancillary qubits in the state $|0\rangle$. Since $f(g)$ can be computed efficiently classically, there exist a efficient quantum circuit realizing $ U_f: |g\rangle|a\rangle\to |g\rangle |f(g)+a\rangle$ for every $k$-bit string $a$ (where $f(g)+a$ is computed modulo 2) so that in particular $|g\rangle|0\rangle$ is mapped to $|g\rangle |f(g)\rangle$. We now apply the single-qubit gate diag$(1, \omega_j)$ to qubit $j$ in the second register, for all $j$. Using (\ref{xi_f(g)}) this maps  \be |g\rangle |f(g)\rangle\to \xi(g) |g\rangle|f(g)\rangle.\ee Finally, applying the inverse of $U_f$ maps $\xi(g) |g\rangle|f(g)\rangle \to \xi(g) |g\rangle|0\rangle$. This yields the desired efficient circuit that realizes $U_{\xi}$.

\section{Proof of theorem \ref{thm_inverse_orthogonal}}\label{sect_app_inverse_orthogonal}

\subsection{Proof of theorem \ref{thm_inverse_orthogonal}a}

Let $h^1, \dots, h^n$ represent a generating set of $H$. The orthogonal subgroup $H^{\perp}$ consists of all $g\in G$ satisfying $\chi_{h^i}(g)=1$ for all $i$. Let $U(1)$ denote the (multiplicative) Abelian group of all complex phases, where 1 is the neutral element. Let $U(1)^n$ be the group of all $n$-tuples of complex phases where the group operation is componentwise multiplication; the neutral element is $(1, \dots, 1)$. For every $g\in G$ we extend the domain of the function $\chi_g$ to the set of all $m$-tuples of integers i.e. if $k=k_1\cdots k_m\in \mathbf{Z}^m$ with $k_i$ integer, one has
\be\label{tilde_chi}
\begin{array}{c}
\chi_{{  g}}:   k\   \to \   \exp 2\pi i\left( \frac{g_1k_1}{d_1}+\dots +\frac{g_mk_m}{d_m} \right),
\end{array}
\ee where, with a slight abuse of notation, we denote the function (\ref{tilde_chi}) by $\chi_g$ as well. We now introduce the following maps.

\vspace{2mm}

$\eta:  G\to U(1)^n$ maps $g\to (\chi_{h^1}(g), \dots, \chi_{h^n}(g))$

\vspace{1mm}

$\varphi: \mathbf{Z}^m\to U(1)^n$ maps  $k \to (\chi_{h^1}(k), \dots, \chi_{h^n}(k))$.

\vspace{1mm}

$\pi : \mathbf{Z}^m\to G$ maps $k\to (k_1 \mbox{ mod } d_1, \dots, k_m\mbox{ mod } d_m)$.

\vspace{2mm}

\noindent Remark that $\pi$ is the natural ``projection'' of $\mathbf{Z}^m$ onto $G$. Every $g\in G$ can naturally be interpreted as an $m$-tuple of integers; we denote this tuple by $\tilde g$ if we want to emphasize this way of interpreting $g$. Note that $\pi(\tilde g)=g$.
\begin{lem}\label{thm_orthogonal}
The maps $\eta, \varphi$ and $\pi$ are group homomorphisms. Furthermore $\varphi=\eta\cdot \pi$. Finally, $H^{\perp}$ coincides with the kernel of $\eta$.
\end{lem}
{\it Proof:} all properties can be verified by a straightforward application of definitions, except the identity $\varphi=\eta\cdot \pi$ which we address next. Let $k=k_1\cdots k_m$ be any $m$-tuple of integers and denote $\pi(k)=l_1\cdots l_m$ with $l_i\in\mathbf{Z}_{d_i}$. Then (by the definition of $x$ mod $y$) there exist integers $c_i$ such that $k_i= l_i + c_i d_i$. Insertion in (\ref{tilde_chi}) shows that $\chi_g(k) = \chi_g(\pi(k))$ for every $g\in G$. It follows that $\varphi(k)=\eta(\pi(k))$. \finpr

To prove theorem \ref{thm_inverse_orthogonal} our goal is to compute a generating set of the kernel of $\eta$. Our approach will be to first compute a generating set of the kernel of $\varphi$, and then to map this solution to a generating set of ker $\eta$ via the projection $\pi$. Remark that $k\in$ ker $\varphi$ iff
\be\label{orthogonal_diophantine1} a_{i1} k_1  + \cdots + a_{im} k_m   \equiv 0 \mod \mathfrak{g};\quad \mbox{where } a_{ij}:= \mathfrak{g}h_j^i / d_j. \ee Remark that $d_i$ divides $\mathfrak{g}$ so that the $a_{ij}$ are integers. Equivalent to (\ref{orthogonal_diophantine1}), there exists a tuple of  integers $s = s_1\dots s_n$ such that
$\label{orthogonal_diophantine2} \sum a_{ij} k_j  + \mathfrak{g} s_i  = 0$.  This is a system of linear diophantine equations in the unknowns $(k, s)$. Owing to theorem \ref{thm_diophantine} we can find a generating set of solutions $(k^1, s^1), \dots (k^r,  s^r)$ in polylog$(\mathfrak{g})$ time. It then follows that the $k^i$ form a generating of solutions to the equations (\ref{orthogonal_diophantine1}) and thus are a generating set of ker $\varphi$.

Define $g^i:=\pi(k^i)$ for every $i$. Note that each $g^i$ can be computed efficiently since this computation involves taking $m$ remainders. We now claim that the $g^i$ generate ker $\eta$. First, we have $g^i\in$ ker $\eta$: indeed by construction $k^i\in $ ker $\varphi$ and owing to lemma \ref{thm_orthogonal} we have $\varphi = \eta\cdot\pi$, implying that $\pi(k^i)\in $ ker $\eta$. Second, we show that every element in the kernel of  $\eta$ can be written as an integer linear combination of the $g^i$. Consider an arbitrary $g\in $ ker $\eta$. Interpreting $g$ as an element in $\mathbf{Z}^m$ we thus have $\tilde g\in$ ker $\varphi$. Since the $k^i$ generate ker $\varphi$, there exist integers $c_i$ such that $\tilde g = \sum c_i k^i$. Since $\pi$ is a homomorphism, it follows that $\pi(\tilde g) = \sum c_i \pi(k^i)$. Using that $\pi(\tilde g)=g$ and $g^i=\pi(k^i)$ we conclude that $g = \sum c_i g^i$ as desired. This proves theorem \ref{thm_inverse_orthogonal}a.

\subsection{Proof of theorem \ref{thm_inverse_orthogonal}b}

The proof method is analogous to above (although the argument is simpler since we do not need to compute a generating set of solutions but merely a single solution) and we just give a sketch. Using the notations introduced above, our goal is to compute a $g\in G$ satisfying \be \eta(g) = (e^{2\pi i s_1/\mathfrak{g}}, \dots, e^{2\pi i s_n/\mathfrak{g}})=: v.\ee
Analogous to above (see (\ref{orthogonal_diophantine1})) we can write down a system of diophantine equations which has a solution iff there exists $k\in \mathbf{Z}^m$ satisfying $\varphi(k)=v$. Deciding whether a solution exists can be done efficiently by virtue of theorem \ref{thm_diophantine}. If a solution exists, owing to the same theorem such a solution $k$ can be computed efficiently. We then compute $g:=\pi(k)$ which can also be done efficiently. Since $\varphi = \eta\cdot \pi$ it follows that $\eta(g) = v$ as desired.

\subsection{Proof of theorem \ref{thm_inverse_orthogonal}c}

\begin{lem}
Let $\alpha$ be an automorphism of $G$ with matrix representation $A$. Then there exists a unique matrix $B$ where each element in the $i$-th row of $B$ belongs to $\mathbf{Z}_{d_i}$ and such that the following identities hold:
\be d_j B_{ij}&\equiv& 0 \mod d_i \quad\mbox{ for every }i, j\label{conditions_endo}\\ \sum_{j=1}^m B_{ij}A_{ji}&\equiv &1\mod d_i\quad\mbox{ for every }i\label{conditions_inverse1}\\ \sum_{j=1}^m B_{kj}A_{ji}&\equiv& 0\mod d_k \quad\mbox{ for every } k\neq i. \label{conditions_inverse2}\ee
Furthermore this unique solution $B$ is the matrix representation of $\alpha^{-1}$.
\end{lem}
{\it Proof: } to show existence, it is straightforward to verify (with lemma \ref{thm_matrix_endo}) the matrix representation of $\alpha^{-1}$ satisfies (\ref{conditions_endo})-(\ref{conditions_inverse2}). To prove uniqueness,  (\ref{conditions_endo}) and lemma \ref{thm_matrix_endo} imply that there exists an endomorphism $\beta$ of $G$ with matrix representation $B$. Conditions (\ref{conditions_inverse1})-(\ref{conditions_inverse2}) imply that $\beta\alpha(e^i)=e^i$ for every $i$. Since $\alpha$ and $\beta$ are endomorphisms, it follows that $\beta\alpha(g)=g$ for every $g\in G$. But then $\beta$ must be identical to the inverse of $\alpha$, which always exists and which is unique. But then also $B$ is unique. \finpr

Our proof of theorem \ref{thm_inverse_orthogonal}c now proceeds as follows. For every $m\times m$ integer matrix $X$, let $\Pi(X)$ denote the matrix obtained by replacing the entry $X_{ij}$ by $X_{ij}$ mod $d_i$. We now first compute an integer matrix $B$ satisfying (\ref{conditions_endo})-(\ref{conditions_inverse2}),  thus for the moment ignoring the constraint that the $i$-th row of $B$ should only contain elements of $\mathbf{Z}_{d_i}$. Computing such an integer matrix solution can be done efficiently by reducing the problem to a system of diophantine equations (by introducing extra integers variables as before). Given an integer solution $B$, we compute $B':=\Pi(B)$; this can be done efficiently. Remark that every element in the $i$-th row of $B'$ does belong to $\mathbf{Z}_{d_i}$. We now claim that $B'$ is still a solution to  (\ref{conditions_endo})-(\ref{conditions_inverse2}). By definition of the map $\Pi$, for every $(i,j)$ there exists an integer $C_{ij}$ which is an integer multiple of $d_i$,  such that $B_{ij}' = B_{ij} + C_{ij}$. Since each $C_{ij}$ is an integer multiple of $d_i$, one has
\be d_j C_{ij}&\equiv& 0 \mod d_i \quad\mbox{ for every }i, j\nonumber\\
\sum_{j=1}^m C_{kj}A_{ji}&\equiv& 0\mod d_k \quad\mbox{ for every } k. \label{conditions_inverse_C}\ee
Using (\ref{conditions_inverse_C}) in combination with the property that $B$ solves (\ref{conditions_endo})-(\ref{conditions_inverse2}), it follows that $B'$ is a solution to these equations as well. Thus $B'$ is the matrix representation of $\alpha^{-1}$.

\section{Proof of theorem \ref{thm_M_state1}}\label{sect_app_M_states}

The normalized sum $\rho := (1/|{\cal G}|) \sum U$ over all $U\in {\cal G}$ is the orthogonal projector on the space of all +1 common eigensvectors of ${\cal G}$ (this property holds for arbitrary unitary groups, see e.g. \cite{Va11}). Since $|\psi\rangle$ is the unique joint eigenvector with eigenvalue 1 by assumption, we thus have $\rho = |\psi\rangle\langle\psi|$. Consider an arbitrary $g$ such that $\langle g|\psi\rangle\neq 0$.  Then \be\label{rho_psi} |\psi\rangle \propto |\psi\rangle\langle\psi|{  g}\rangle = \frac{1}{|\cal G|}\sum U|{  g}\rangle.\ee This shows that all amplitudes $\langle {  h}|\psi\rangle$ are zero whenever ${  h}$ lies \emph{outside} the orbit $O_g$. Now consider an arbitrary $h\in O_g$. Then, by definition of $O_g$, there exist $U\in {\cal G}$ satisfying $U|g\rangle = d_{g}|h\rangle$ where $d_g$ is the diagonal entry of $D$ on position $g$.  Remark that $d_g$ is a complex phase since $U$ is unitary. Using that $U^{\dagger}|\psi\rangle = |\psi\rangle$ it follows that $\langle h|\psi\rangle = d_g \langle {  g}|\psi\rangle.$ This proves that all amplitudes $\langle h|\psi\rangle$ with $h\in O_g$ are equal in modulus. This shows (a) and (b).

Next we prove (c). Suppose first that $O_g=S$ and consider a generator $V_i$. Then there exists a complex phase $\xi$ such that $V_i|g\rangle = \xi|g\rangle$. It follows that \be \label{xi}\langle\psi |V_i|g\rangle = \xi \langle\psi|g\rangle.\ee Using that $\langle\psi |V_i = \langle\psi|$ since $V_i$ belongs to ${\cal G}$, one finds that $\langle\psi|g\rangle = \xi \langle\psi|g\rangle$. Since $O_g= S$ and $g\in O_g$, it follows that $\langle\psi|g\rangle\neq 0$. With (\ref{xi}) this shows that $\xi=1$. Conversely, suppose that $g$ satisfies $V_i|g\rangle = |g\rangle$ for every $i$. As a consequence, $V|g\rangle = |g\rangle$ for every $V\in {\cal G}_g$. Using that $\rho = |\psi\rangle\langle\psi|$ yields \be |\langle \psi|g\rangle|^2 &=& \frac{1}{|{\cal G}|} \sum_{U\in {\cal G}} \langle g|U|g\rangle = \frac{1}{|{\cal G}|} \sum_{U\in {\cal G}_g} \langle g|U|g\rangle= \frac{|{\cal G}_x|}{|{\cal G}|}\neq 0.\ee In the second identity we have used that $\langle g|U|g\rangle=0$ if $U\notin {\cal G}_x$ since $U$ is monomial; in the third identity we used that $U|g\rangle = |g\rangle$ for every $U\in {\cal G}_x$. Finally we used that $|{\cal G}_g|\neq 0$ since $I\in {\cal G}_g$.

\section{The quadratic functions (\ref{ex_quadratic_3})}\label{sect_app_quadratic}

We prove that  the function (\ref{ex_quadratic_3}), denoted here by $\xi$, is quadratic. Consider $x, y\in\mathbf{Z}_d$. Within the present section, we denote by $x+y$ the sum of these elements over the integers, whereas $x\oplus y$ denotes the sum in $\mathbf{Z}_d$. With this notation, one has \be x\oplus y &=& x+y\mbox{ mod } d \quad\mbox{ and }\quad q_a(x)= ax( x + d) \label{oplus}\ee It follows from the left-hand equation in (\ref{oplus}) that there exists an integer $b$ such that $x\oplus y = x+y +bd. $ Consequently,
\be q_a(x\oplus y) &=& q_a(x+y+qd)= q_a(x) +   q_a(y) + 2axy + a c\label{oplus2}\\ \mbox{ where } c &:=& 2xbd + 2ybd + d^2 b(b+ 1).\ee Remark that $c$ is an integer multiple of $2d$. Indeed, both terms $2xbd$ and   $2ybd$ are obviously multiples of 2d; furthermore $b(b+1)$ is even for every integer $b$, so that $d^2 b(b+ 1)$ is also a multiple of $2d$. It follows that $\exp[i\pi ac/d]=1$ and thus with (\ref{oplus2}) one finds \be \xi(x\oplus y) &=& e^{i\pi [q_a(x) +   q_a(y) + 2axy + a c]/d}\nonumber \\ &=& e^{i\pi q_a(x)/d} e^{\pi i   q_a(y)/d} e^{2\pi iaxy/d} e^{\pi i a c/d} = \xi(x)\xi(y) B(x, y),\label{last}\ee with $B(x, y):= \exp[2\pi i a xy/d]$. Since the function $B$ is bilinear, this shows that $\xi$ is quadratic.


\begin{thebibliography}{99}
\bibitem{Foot5}
Since $\mathfrak{g}=d_1\cdots d_m$, one has $\log_2 \mathfrak{g} = \sum_{i=1}^m \log_2 d_i$. Since $d_i\geq 2$ one has $\log_2 d_i\geq 1$. This implies that $\log_2 \mathfrak{g}\geq m$.
\bibitem{Foot9}
Here $cx$ is the element in $\mathbf{Z}_{d_2}$ given by $c+\cdots +c$ with $x$ summands.
\end{thebibliography}
\end{document}